\documentclass[preprint]{elsarticle}
\usepackage{graphicx}
\usepackage{amssymb,amsmath}
\usepackage{subfig}
\newcommand{\nc}{\newcommand}
\nc{\rnc}{\renewcommand}
\nc{\x}[1]{\mbox{#1}}           
\nc{\hs}[1]{\hspace*{#1}}
\nc{\vs}[1]{\vspace*{#1}}
\rnc{\theta}{\vartheta}
\rnc{\rho}{\varrho}
\rnc{\epsilon}{\varepsilon}
\nc{\dd}{{\mathrm{d}}}
\nc{\ii}{{\mathrm{i}}}
\nc{\ee}{{\mathrm{e}}}
\nc{\fehat}{{\mms{\hat f}_{\mathrm e}}}
\nc{\mm}[1]{\mathbf{#1}}
\nc{\mms}[1]{\boldsymbol{#1}}
\nc{\suml}[2]{\sum \limits_{#1}^{#2}}
\nc{\intl}[2]{\int \limits_{#1}^{#2}}
\rnc{\matrix}[2]{\left[\!\!\begin{array}{#1} #2\end{array}\!\!\right]}
\rnc{\vector}[1]{\matrix{c}{#1}}
\nc{\absphase}{\theta}
\nc{\inv}{^{-1}}
\nc{\tra}{^{\mathrm T}}
\nc{\herm}{^{\mathrm H}}
\nc{\sgn}{\mathrm{sgn}}
\nc{\diag}{\mbox{\bf diag}}
\nc{\eo}{EO}
\nc{\nh}{{N_{\mathrm h}}}
\nc{\ncp}{{N_{\mathrm{CP}}}}
\nc{\nnl}{{N_{\mathrm{N}}}}
\nc{\nll}{{N_{\mathrm{L}}}}
\nc{\ndim}{{N_{\mathrm{dof}}}}
\nc{\nnlred}{{N_{\mathrm{nl,red}}}}
\nc{\myquote}[1]{`#1'}
\nc{\kreis}[1]{\mbox{\mbox{\text{\Large{\ensuremath{\bigcirc}}}}\hspace{-2.4ex}#1\hspace{1.2ex}}}
\nc{\prob}{\operatorname{Prob}}
\nc{\fnp}{N}
\nc{\fnpopt}{N^{\rm{opt}}}
\nc{\fnpoptc}{N^{\rm{opt}}_{\rm{c}}}
\nc{\fnpoptuc}{N^{\rm{opt}}_{\rm{uc}}}
\nc{\ares}{a^{\rm{res}}}
\nc{\astick}{a^{\rm{res}}_{\rm{stick}}}
\nc{\amin}{a^{\mathrm{opt}}_{\mathrm{c}}}
\nc{\fres}{f^{\rm{res}}}
\nc{\areference}{a^{\rm{ref}}}
\nc{\argm}{\operatorname{arg}}
\nc{\kn}{{k_{\mathrm n}}}
\nc{\knl}{{k_{\mathrm{nl}}}}
\nc{\kt}{{k_{\mathrm t}}}
\nc{\ommod}{{\omega_0}}
\nc{\dmod}{D}
\rnc{\Re}[1]{\operatorname{Re}\lbrace #1 \rbrace}
\rnc{\Im}[1]{\operatorname{Im}\lbrace #1 \rbrace}
\nc{\prog}[1]{{\sf{#1}}}
\nc{\matlab}{\prog{Matlab}}
\nc{\ie}{i.\,e.\,}
\nc{\eg}{e.\,g.\,}
\nc{\cf}{cf.\,}
\nc{\etal}{et~al.\,}
\nc{\z}[2]{\x{\sc #1}~{\x{\cite{#2}}}}
\nc{\zo}[1]{\x{\cite{#1}}}
\nc{\fabstand}{\,}
\nc{\fp}{\fabstand.}
\nc{\fk}{\fabstand,}
\nc{\tab}[5][tbh]{\begin{table}[#1]\centering\caption{#4\label{tab:#5}}\begin{tabular}{#2}\hline #3 \\ \hline\end{tabular}\end{table}}
\newcommand{\fss}[4][tbh]{%
 \begin{figure}[#1]
 \centering
 \if@draft
  \framebox[150mm]{\raisebox{0mm}[25mm][25mm]{\texttt{#2}}}
 \else
  \includegraphics[scale=#4]{#2}
 \fi
 \caption{#3}
 \label{fig:#2}
\end{figure}
}
\newcommand{\fs}[4][tbh]{%
 \begin{figure}[#1]
 \centering
 \if@draft
  \framebox[150mm]{\raisebox{0mm}[25mm][25mm]{\texttt{#2}}}
 \else
  \includegraphics[width=#4\textwidth]{#2}
 \fi
 \caption{#3}
 \label{fig:#2}
\end{figure}
}
\newcommand{\myf}[8][tbh]{%
    \begin{figure}[#1]
        \centering
        \subfloat[#4 \label{fig:#2}]{
            \includegraphics[width=#6\textwidth]{#2}
        }
        \hspace{0.25cm}
        \subfloat[#5 \label{fig:#3}]{
            \includegraphics[width=#7\textwidth]{#3}
        }
        \caption{#8}
    \end{figure}
}

\nc{\e}[2]{\begin{equation} #1 \label {eq:#2} \end{equation}}
\nc{\ea}[2]{
\begin{eqnarray}
#1 \label {eq:#2} \end{eqnarray}}
\nc{\eal}[3][0.0ex]{
\begin{samepage}
\begin{eqnarray*}
#2
\end{eqnarray*}
\nopagebreak[4] \vs{#1} \nopagebreak[4]\vs{-2ex} \nopagebreak[4]
\begin{eqnarray}
\label {eq:#3}
\end{eqnarray}
\end{samepage}\hs{-0.35em}}
\nc{\g}[1]{{$#1$}}
\nc{\gehg}[3]{\x{${#1}= #2 ~ {\rm{#3}}$}}
\nc{\fref}[1]{{Fig.~\ref{fig:#1}}}
\nc{\frefo}[1]{{\ref{fig:#1}}}
\nc{\frefoo}[1]{{#1}}
\nc{\frefs}[1]{{Figs.~\ref{fig:#1}}}
\nc{\tref}[1]{{Tab.~\ref{tab:#1}}}
\nc{\trefo}[1]{{\ref{tab:#1}}}
\nc{\trefs}[1]{{Tab.~\ref{tab:#1}}}
\nc{\erefn}[1]{{Eq.~(\ref{#1})}}
\nc{\eref}[1]{{Eq.~(\ref{eq:#1})}}
\nc{\erefo}[1]{(\ref{eq:#1})}
\nc{\erefs}[2]{{Eqs.~(\ref{eq:#1}) and (\ref{eq:#2})}}
\nc{\erefss}[1]{{Eqs.~(\ref{eq:#1})}}
\nc{\sref}[1]{section~\ref{sec:#1}}
\nc{\srefo}[1]{\ref{sec:#1}}
\nc{\srefs}[1]{{sections~\ref{sec:#1}}}
\nc{\aref}[1]{{{appendix~\ref{asec:#1}}}}
\nc{\arefo}[1]{{\ref{asec:#1}}}
\nc{\arefs}[1]{{{appendices~\ref{asec:#1}}}}
\nc{\ssref}[1]{{subsection~\ref{sec:#1}}}
\nc{\ssrefo}[1]{\ref{sec:#1}}
\nc{\ssrefs}[1]{{subsections~\ref{sec:#1}}}

\begin{document}

\begin{frontmatter}

\title{ON THE COMPUTATION OF THE SLOW DYNAMICS OF
NONLINEAR MODES OF MECHANICAL SYSTEMS}

\author[ids]{Malte Krack\corref{cor1}}
\ead{krack@ila.uni-stuttgart.de}
\author[ids]{Lars Panning-von Scheidt}
\author[ids]{J\"org Wallaschek}


\cortext[cor1]{Corresponding author}

\begin{abstract}
\textit{A novel method for the numerical prediction of the slowly
varying dynamics of nonlinear mechanical systems has been developed.
The method is restricted to the regime of an isolated nonlinear mode
and consists of a two-step procedure: In the first step, a
multiharmonic analysis of the autonomous system is performed to
directly compute the amplitude-dependent characteristics of the
considered nonlinear mode. In the second step, these modal
properties are used to construct a two-dimensional reduced order
model (ROM) that facilitates the efficient computation of
steady-state and unsteady dynamics provided that nonlinear modal
interactions are absent.\\
The proposed methodology is applied to several nonlinear mechanical
systems ranging form single degree-of-freedom to Finite Element
models. Unsteady vibration phenomena such as approaching behavior
towards an equilibrium point or limit cylces, and resonance passages
are studied regarding the effect of various nonlinearities such as
cubic springs, unilateral contact and friction. It is found that the
proposed ROM facilitates very fast and accurate analysis of the slow
dynamics of nonlinear systems. Moreover the ROM concept offers a
huge parameter space including additional linear damping, stiffness
and near-resonant forcing.}
\end{abstract}


\end{frontmatter}

\section*{Nomenclature}
\begin{eqnarray}
\nonumber a && \text{Modal amplitude}\\
\nonumber \mms f && \text{Nonlinear force vector}\\
\nonumber \ii && \text{Imaginary unit}\\
\nonumber \mms M, \mms C, \mms K && \text{Mass, damping, stiffness matrices}\\
\nonumber N_{\mathrm h} && \text{Harmonic order of Fourier ansatz}\\
\nonumber t && \text{Time}\\
\nonumber \mms u && \text{Displacement vector}\\
\nonumber \mms u_{\mathrm p} && \text{Periodic form of displacement vector}\\
\nonumber \mms v && \text{Complex eigenvector}\\
\nonumber \lambda, \ommod, \dmod && \text{Eigenvalue, eigenfrequency, damping ratio}\\
\nonumber \Omega && \text{Angular frequency of oscillation}\\
\nonumber \varphi, \varphi_{\mathrm e} && \text{Fast phase, fast phase induced by excitation}\\
\nonumber \mms \Psi_n && \text{N-th harmonic of eigenvector}\\
\nonumber \Theta && \text{Slow phase}\\
\nonumber \absphase && \text{Absolute phase}\\
\nonumber \dot{\left(~\right)}, \ddot{\left(~\right)} && \text{First and second-order derivative with respect to time $t$}\\
\nonumber \bar{\left(~\right)} && \text{Complex conjugate}\\
\nonumber {\left(~\right)}\tra && \text{Transpose}\\
\nonumber {\left(~\right)}\herm && \text{Hermitian transpose}\\
\nonumber \langle\cdot,\cdot\rangle && \text{Inner product}\\
\nonumber\text{DOF} && \text{Degree of freedom}\\
\nonumber\text{FE} && \text{Finite Element}\\
\nonumber\text{HBM} && \text{(High-order) Harmonic Balance Method}\\
\nonumber\text{NMA} && \text{Nonlinear Modal Analysis}\\
\nonumber\text{ODE} && \text{Ordinary Differential Equation}\\
\nonumber\text{ROM} && \text{Reduced Order Model}
\end{eqnarray}

\section{Introduction\label{sec:introduction}}
\subsection{Motivation for reduced order modeling of nonlinear systems}
In many structural dynamic systems, the effect of nonlinearity plays
an important role. In fact, several engineering applications exploit
nonlinear phenomena in order to improve the dynamic behavior of
mechanical structures. Hence, there is a considerable need for
efficient and versatile methods for the dynamic analysis of such
systems.\\
For systems that comprise a large number of degrees of freedom
(DOFs) and exhibit generic nonlinearities, as addressed in this
study, the applicability of analytical methods is typically not
possible and numerical methods have to be employed. The application
of direct solution methods such as time-step integration often
results in high computational costs. Therefore, extensive parametric
studies, sensitivity and uncertainty analyses or design optimization
soon become infeasible in conjunction with the full order model.
Thus, there is a demand for Reduced Order Models (ROM) that are
capable of significantly reducing the computational effort for the
dynamic analysis and retaining the required accuracy of the
predicted results.

\subsection{Existing approaches}
Methods based on the invariant manifold approach
\zo{shaw1993,nayf2000,jian2005,pierre2006,touz2006} are widely-used
for the modal analysis of nonlinear mechanical systems as well as
for the construction of efficient ROMs. The approach is based on the
invariance property of certain periodic orbits of the dynamical
systems, \ie a nonlinear mode is defined as an invariant
relationship (manifold) between several master coordinates and the
remaining coordinates of the system. This manifold is typically
governed by partial differential equations arising from the
substitution of the manifold into the state form of the equations of
motion. The invariant manifold approach was extended to account for
the effect of harmonic excitation \zo{jian2005} and viscous damping
\zo{touz2006}. It has been applied to various problems including
piecewise linear systems \zo{jian2004}, internally resonant
nonlinear modes \zo{pierre2006} and generic nonlinearly damped
systems
\zo{rens2013}.\\
A straight-forward ROM strategy consists of constraining the system
dynamics to the computed invariant manifold. This strategy is
capable of drastically reducing the dimensionality of the problem to
only a few master coordinates, while providing excellent accuracy
for steady-state as well as unsteady dynamic predictions. One
drawback is, however, the huge effort for the computation of the
invariant manifolds. These computations can involve several
thousands of nonlinearly coupled algebraic equations \zo{jian2005}.
Moreover, the development of numerically robust algorithms for the
treatment of generic, in particular non-conservative nonlinearities,
seems to be an unresolved problem, see \eg \zo{rens2013}.
Furthermore, since the time-dependency is lost in the problem
definition, the characteristic frequencies cannot directly be
obtained from the computed manifold, but has to be identified from
simulation results \zo{jian2004,rens2013}. Finally, the parameter
space of the ROM based on the invariant manifold approach is
typically limited. For example, harmonic excitation and viscous
damping are generally considered in the manifold computation step
\zo{jian2005,touz2006} so that even slight modification of these
parameters would require
the re-computation of the manifold.\\\\
Another category of methods for the determination of modal
properties of nonlinear systems can be classified as nonlinear
system identification (NSI) approaches
\zo{chon2000a,gibe2003,kers2005,lee2010}. Response data, obtained
either by simulation or measurement, is gathered and modal
properties are identified by fitting original response data to data
from nonlinear modal synthesis. The weak point of this strategy is
clearly its signal-dependent nature and the further effort required
to obtain the response data. One of the main benefits of this method
is that no model is required for the nonlinearities which enables
broad applicability. Modal properties identified with NSI methods
can be easily employed to feed the parameters of a nonlinear ROM
\zo{chon2000a,gibe2003}.\\\\
Harmonic Balance approaches are widely used for the nonlinear modal
analysis of conservative mechanical systems
\zo{leun1992,ribe2000,coch2009}. These methods are generally known
to be well-suited for the analysis of strongly nonlinear systems
with a large number of DOFs. Recently, \z{Laxalde and
Thouverez}{laxa2009} extended the Harmonic Balance Method to the
approximate modal analysis of dissipative systems by introducing a
complex eigenfrequency in the Fourier ansatz. \z{Krack
\etal}{krac2013a,krac2013d} applied this approach to various
nonlinear mechanical systems and significantly improved the
numerical performance of the analysis. This frequency-domain modal
analysis concept allows for the direct calculation of iso-energy
orbits on the invariant manifold as well as eigenfrequency and modal
damping ratio of the nonlinear system. The modal properties have
also been exploited in a ROM formulation for the prediction of
steady-state vibrations \zo{laxa2009,krac2013a,krac2013d}. \z{Krack
\etal}{krac2013a,krac2013d} also investigated the limitations of the
ROM and demonstrated that the validity of this approach is
restricted to those regimes in which the energy is confined to a
single nonlinear mode. This finding is completely in line with the
work of \z{Blanc \etal}{blan2013}, who concluded that the absence of
nonlinear modal interactions represents an intrinsic limitation to
ROMs based on the invariant manifold concept.

\subsection{Need for research regarding approaches for unsteady dynamics}
For various engineering applications, it is important to assess the
transient dynamics induced `on the way to' the operating point, if
it exists. Most of the above mentioned ROM concepts are, however,
designed to predict the steady-state vibration behavior of nonlinear
systems. Among the few existing approximation methods for the slow
dynamics of nonlinear systems, complexification-averaging is
probably the most commonly used technique
\zo{mane2001,lee2005,vaka2008b,vaka2008}. This technique is,
however, typically applied to obtain closed-form analytical
solutions for systems featuring polynomial nonlinearities and only a
small number of degrees of freedom.\\
In order to obtain a concept for generic nonlinear systems, this
study builds on the broadly applicable multi-harmonic nonlinear
modal analysis technique developed in \zo{laxa2009,krac2013a}. The
original problem and considered dynamic regime is described in
\sref{problem_statement}. The generalized Fourier-Galerkin method is
briefly revisited in \sref{nma_revisited}. For the first time, the
averaging technique is applied to the thus defined nonlinear modes
in \sref{slow_dynamics} to obtain approximations for the slow
dynamics on the invariant manifold of an isolated nonlinear mode. In
order to assess the performance and validity of the proposed
methodology, several nonlinear dynamical problems are addressed in
\sref{numerical_examples}. Conclusions are drawn in
\sref{conclusions}.

\section{Original problem and considered dynamic regime
\label{sec:problem_statement}}%
Consider a discrete, time-invariant, mechanical system whose
dynamics is governed by a second-order ordinary differential
equation (ODE),
\e{\mms{M}\ddot{\mms{u}}(t)+
\mms{f}\left(\mms{u}(t),\dot{\mms{u}}(t)\right) = -\mms{\tilde
K}\mms u - \mms {\tilde C}\mms{\dot u} +
\fehat~\frac{\ee^{\ii\phi_{\mathrm e}(t)}+\ee^{-\ii\phi_{\mathrm
e}(t)}}{2}\fp}{eqm_perturbed}
Herein, \g{\mms M = \mms M\tra>0} is the real, symmetric, positive
definite mass matrix and \g{\mms u(t)} is the vector of generalized
coordinates. The operator \g{\mms f} comprises nonlinear functions
in displacement and velocity, and does not explicitly depend on time
so that parametric excitation is excluded. \g{\mms{\tilde K}} and
\g{\mms{\tilde C}} are symmetric linear stiffness and damping terms,
respectively. \g{\fehat} and \g{\phi_{\mathrm e}(t)} represent
amplitude vector and phase function of the external forcing.\\
Throughout this study, the considered dynamics are limited to the
regime where nonlinear modal interactions are absent and the energy
is confined in an isolated nonlinear mode. It is further assumed
that the linear stiffness and damping terms on the right hand side
of \eref{eqm_perturbed} are weak in the sense that they do not
significantly deteriorate the geometry of the invariant manifold
associated with the considered mode. A transient forcing can be
taken into account in \eref{eqm_perturbed}, but the phase function
\g{\phi_{\mathrm e}(t)} is assumed to exhibit a dominant
instantaneous frequency near the fundamental resonance of the
considered mode. In accordance with the deformation-at-resonance
property of nonlinear modes \zo{vaka2008}, it is therefore
reasonable to assume that the external forcing only controls the
amplitude but not the shape of the nonlinear mode. The damping
matrix \g{\mms{\tilde C}} may be indefinite so that self-excitation
is possible in the absence of external forcing. However, it is
assumed that the damping matrix is positive if external forcing is
present.\\
With these assumptions, it is possible to neglect the weak stiffness
and damping terms on the right hand side of \eref{eqm_perturbed} for
modal analysis step described in \sref{nma_revisited} and to
re-introduce these effects in ROM formulation in
\sref{slow_dynamics}. The motivation for this is to equip the ROM
with a high-dimensional parameter space, since the parameters
associated with external forcing, damping and stiffness are
retained. This makes the proposed approach particularly attractive
for exhaustive parametric studies. Furthermore, the modal analysis
is simplified in this approach compared to approaches where forcing
is taken into account in the modal analysis step, since the
dimension of the phase space is reduced \zo{jian2005,touz2006,pierre2006}.\\
Instead of computing the nonlinear modes directly for the problem in
\eref{eqm_perturbed}, they are therefore computed for the autonomous
surrogate problem \eref{eqm_autonomous},
\e{\mms{M}\ddot{\mms{u}}(t)+
\mms{f}\left(\mms{u}(t),\dot{\mms{u}}(t)\right)
=\mms{0}\fp}{eqm_autonomous}
In this study, the slow dynamics, \ie the unsteady dynamics with a
fast oscillation with slowly varying amplitude and phase is
addressed.

\section{Formulation and solution of the complex eigenvalue problem\label{sec:nma_revisited}}
As in a linear modal analysis, the problem \eref{eqm_autonomous} is
solved using an exponential ansatz. In contrast to a linear system,
however, the notion of eigenvectors is generalized so that the
\textit{eigenvectors} can comprise multi-harmonic content. Hence,
\g{\mms u(t)} is expressed in terms of the modal amplitude \g{a} and
the complex components \g{\mms\Psi_n},
\e{\mms u(t) = a\Re{\suml{n=0}{\nh}\mms\Psi_n\ee^{n\lambda
t}}\fp}{fourier_ansatz}
Note that the generalized Fourier series is truncated to the
harmonic order \g{\nh}. For a nonlinear system, all harmonics can be
non-zero in general, including the zeroth-order term \g{\mms\Psi_0}
which corresponds to a static offset. The formulation of
\eref{fourier_ansatz} implies that the complex fundamental
eigenfrequency \g{\lambda} is the same for all harmonics. The
undamped eigenfrequency \g{\ommod} and the modal damping ratio
\g{\dmod} are related to \g{\lambda} as follows:
\e{\lambda =
-\dmod\ommod+\ii\ommod\sqrt{1-\dmod^2}\fp}{eigenfrequency}
%
The set \g{\lbrace\lambda,\mms\Psi_0,\cdots\mms\Psi_{\nh}\rbrace}
consisting of complex
eigenfrequency and harmonic components is denoted eigenpair.\\
The eigenproblem is solved in the frequency domain. Therefore, the
ansatz in \eref{fourier_ansatz} is substituted into
\eref{eqm_autonomous} and subsequent Fourier-Galerkin projection
yields a nonlinear system of algebraic equations. Similar to the
linear case, amplitude and phase normalization constraints are
imposed to make the number of equations equal to the number of
unknowns. The resulting \textit{complex eigenproblem} can be stated
as:
\ea{\nonumber\text{solve} & (n\lambda)^2\mms M\mms\Psi_n a +
\langle\mms f(\mms{u}_{\mathrm p},\mms{\dot{u}}_{\mathrm
p}),\ee^{\ii n\ommod t}\rangle =
\mms 0\,,\,\, n=0\cdots\nh\\
\nonumber\text{subject to} & \underbrace{\mms\Psi_1\herm\mms
M\mms\Psi_1=1}_{\text{amplitude norm.}}\,,\,\,
\underbrace{\Re{\mms t\herm\mms\Psi_1}=0}_{\text{phase norm.}}\\
\text{with respect to} &
\lbrace\lambda,\mms\Psi_0,\cdots\mms\Psi_{\nh}\rbrace\fp}{complex_evp}
The constant vector \g{\mms t} is used here to constrain the phase
of the fundamental harmonic of the eigenvector \g{\mms\Psi_1}. The
amplitude normalization to the mass matrix makes the formulation
consistent with the classical formulation of the linear modal
analysis.\\
An important aspect of the formulation in \eref{complex_evp} is that
the modal analysis involves the periodic forms of displacement
\g{\mms{u}_{\mathrm p}}, velocity \g{\mms{\dot u}_{\mathrm p}}
instead of the pseudo-periodic ones in \eref{fourier_ansatz} as in
\zo{laxa2009,krac2013a}. These periodic forms are defined as
follows:
\e{\mms{u}_{\mathrm p} = a\Re{\suml{n=0}{\nh}\mms\Psi_n\ee^{\ii
n\ommod t}}\fk\quad \mms{\dot u}_{\mathrm p} =
a\Re{\suml{n=0}{\nh}\ii n\ommod\mms\Psi_n\ee^{\ii n\ommod
t}}\fk}{periodic_eigenvector}
and the associated inner product
\e{\langle c(\ommod t),d(\ommod t)\rangle =
\frac{1}{2\pi}\intl{(2\pi)}{}{c(\ommod t)\overline{d}(\ommod
t)}\dd\ommod t\fp}{inner_product}
The strategy of using periodic formulations inherently allows for
the direct calculation of periodic orbits and their associated
harmonic decomposition. The computed nonlinear modes are therefore
consistent with steady-state conditions. Since we are interested in
dynamics with slowly increasing as well as decreasing energy, this
approach gives rise to results that are centered with respect to
energy.\\
The generalized Fourier coefficients \g{\langle\mms
f(\mms{u}_{\mathrm p},\mms{\dot{u}}_{\mathrm p}),\ee^{\ii n\ommod
t}\rangle} of the nonlinear forces in \eref{complex_evp} can
typically not be expressed in closed-form and in general has to be
evaluated numerically. Using periodic forms in the evaluation of the
generalized Fourier coefficients facilitates the application of
existing (high-order) harmonic balance formulations which require
this periodicity. Available formulations exist for various
conservative and non-conservative nonlinear systems (see \eg
\zo{came1989,guil1998,coch2009,krac2013b}) and shall not be
repeated at this point.\\
The frequency-domain solution of the nonlinear eigenproblem can be
regarded as a straight-forward extension of the linear eigenproblem.
The governing algebraic system of equations can be solved using \eg
a Newton-Raphson method. Typically, the linearized eigenpair
represents a valid and often good initial guess for small modal
amplitudes. The efficiency of the solution process can be enhanced
by using analytical gradients as proposed in \zo{krac2013a}. An
exact condensation of the algebraic system of equations can
significantly reduce the required computational effort in case of
sparse nonlinear operators \zo{krac2013a} present \eg in jointed
structures. The solution of \eref{complex_evp} has to be carried out
for each nonlinear mode of interest in the relevant modal amplitude
range.\\
The range of validity of the proposed reduction to a single
nonlinear mode is limited to the regime where the considered
nonlinear mode is stable and unique. Nonlinear modal interactions
can lead to folds and bifurcations of nonlinear modes
\zo{krac2013a,kers2009,coch2009,ribe2000}. Continuation with respect
to the modal amplitude in conjunction with stability and bifurcation
analysis can reveal these phenomena. Since the problem is formulated
in the frequency domain, stability can be analyzed using Hill's
theory \zo{groll2001b,laza2010}. Alternatively, Floquet theory can
be directly applied by computing and investigating the eigenvalues
of the monodromy matrix associated with the periodic orbits
\zo{sund1997,jian2004,lee2005}.

\subsection{Definition of invariant manifolds}
It should be noted that the phase normalization was only imposed in
\eref{complex_evp} in order to make the number of equations equal to
the number of unknowns. In the autonomous case, the phase is
generally arbitrary and an absolute phase \g{\absphase} can be
introduced. Hence, the eigensolution can be generally written as
\ea{\nonumber\mms u = \mms P(a,\absphase) := a\frac{\mms
v+\mms{\overline v}}{2}\,,\quad \mms{\dot u} = \mms
Q(a,\absphase) := \mms{\dot P}(a,\absphase)\fk\\
\text{with}\quad \mms v(a,\absphase) =
\suml{n=0}{\nh}\mms\Psi_n(a)\ee^{\ii n\absphase}\fp}{manifold}
Clearly, \eref{manifold} defines a two-dimensional invariant
manifold in phase space, consistent with the invariant manifold
concept developed by \textsc{Shaw and Pierre}\zo{shaw1993}. The
manifold is described in a polar coordinate system with the modal
amplitude \g{a(t)} and the absolute phase \g{\absphase(t)} which
both can be arbitrary functions in time $t$. It should be remarked
that the term \g{\mms Q} is only defined here for the sake of
completeness of the phase space and the time derivative is not
carried out at this point.

\section{Prediction of the slow dynamics of the nonlinear mode\label{sec:slow_dynamics}}
In accordance with the single nonlinear mode theory \zo{szem1979},
the ROM in this study is restricted to the regime where nonlinear
modal interactions are absent. If this assumption does not hold, a
higher-dimensional ROM would have to be developed \zo{pierre2006}.
It should be recalled that the weak damping and stiffness terms, as
well as the fundamental near-resonant forcing in
\eref{eqm_perturbed} have to remain small enough in order to achieve
accurate predictions with the ROM based on the nonlinear modes of
the surrogate problem \eref{eqm_autonomous} in which these effects
are neglected. Significant deviations are expected for fast
transient phenomena, strong variation of the eigenvector due to the
additional damping and stiffness terms and in situations where the
system is driven
into energy regimes featuring nonlinear modal interactions.\\
In order to derive the ODEs governing the slow dynamics on the
manifold, the complexification-averaging technique \zo{mane2001} is
employed. Therefore, a coordinate transform is defined,
\e{\mms u := a\frac{\mms v(a,\absphase) + \mms{\overline
v}(a,\absphase)}{2}\fk\quad \mms{\dot u} := a\ii\Omega\frac{\mms
v(a,\absphase) - \mms{\overline
v}(a,\absphase)}{2}\fp}{complexification}
The scalar complex variable conventionally introduced in the
complexification-averaging process is here generalized to the
multiharmonic vector \g{\mms v} as defined in \eref{manifold}. In
\eref{complexification}, \g{\absphase=\absphase(t)} is the time
dependent
phase of the mode and \g{\Omega} is the frequency of oscillation.\\
The key point of the approximation is the decomposition into fast
and slow dynamics, \ie fast and slowly varying components of the
phase \g{\absphase(t)},
\ea{\absphase(t) = \underbrace{\phi(t)}_{\text{fast}} +
\underbrace{\Theta(t)}_{\text{slow}}\fk\\
\Omega:=\dot\phi(t)\fp}{phase}
The difference between the time scale of fast phase \g{\phi(t)} and
slow phase \g{\Theta(t)} is assumed to be prominent, so that it is
valid to approximate the angular frequency \g{\Omega} by the time
derivative of the fast phase.\\
The fast phase \g{\phi(t)} corresponds to the eigenfrequency
\g{\ommod} in the autonomous case, and is assumed to be induced by
excitation in the forced case,
\e{\dot\phi(t):=\begin{cases}\ommod & \text{autonomous
dynamics}\\
\dot\phi_{\mathrm e}(t) & \text{non-autonomous dynamics}
\end{cases}\fp}{fastphase}
Subsequently, the classical averaging process can be carried out. A
detailed derivation is given in \aref{appendix}. The resulting first
order ODEs in \g{a,\Theta} read
\e{\vector{\dot a\\ \dot\Theta}  =
\frac{1}{2\Omega}\vector{-2\tilde\dmod\tilde\ommod\Omega a -
\mms\Psi_1\herm\mms
{\fehat}\sin\Theta\\
\tilde\ommod^2-\Omega^2-\frac{1}{a}\mms\Psi_1\herm\mms
{\fehat}\cos\Theta}\fp}{odeslowflow}
Herein, the modified modal properties are defined as
\e{\quad \tilde\ommod^2 = \ommod^2+\mms\Psi_1\herm\mms{\tilde
K}\mms\Psi_1\fk \quad 2\tilde \dmod\tilde \ommod =
2\dmod\tilde\ommod +\mms\Psi_1\herm\mms {\tilde
C}\mms\Psi_1\fp}{perturbedmodalproperties}
It should be emphasized that the modal properties
\g{\ommod(a),\dmod(a),\mms\Psi_1(a)} depend on the modal amplitude
\g{a}, turning \eref{odeslowflow} into a nonlinear problem.\\
The solution of \eref{perturbedmodalproperties} is particularly
efficient in this formulation for two reasons: (1) The
dimensionality of the problem is at maximum two, (2) the evaluation
of the nonlinear terms in \eref{perturbedmodalproperties} does not
involve the original possibly expensive nonlinear operator \g{\mms
f} but only the readily available amplitude-dependent modal
properties. This procedure therefore combines the highly accurate
multiharmonic analysis of the full system subject to various,
possibly strong nonlinearities, with the small and simple problem in
\eref{perturbedmodalproperties}.\\
In a numerical procedure, the modal properties will only be computed
at discrete amplitude values. Hence, a one-dimensional interpolation
scheme has to be used in order to apply the continuous formulation
in \eref{odeslowflow} to the numerical results of the modal
analysis.\\
Some special cases should be highlighted at this point: Under
\textit{steady-state} conditions \g{\dot a=0=\dot\Theta}, the ODE
system in \eref{odeslowflow} degenerates to an algebraic system of
equations in \g{a,\Theta}, as already presented in \zo{krac2013a}.
Otherwise, \eref{odeslowflow} in conjunction with appropriate
initial conditions \g{a(t=0)=a_0}, \g{\Theta(t=0)=\Theta_0},
represents an initial value problem that governs the slow dynamics
of the considered system. In the \textit{autonomous case}, the
second line of \eref{odeslowflow} yields \g{\Theta=\Theta_0} and the
first line degenerates to \g{\dot a = \tilde\dmod\left(a\right)\tilde\ommod\left(a\right) a}.

\subsection{Projecting the initial state onto the manifold\label{sec:projectstate}}
\fs[t!]{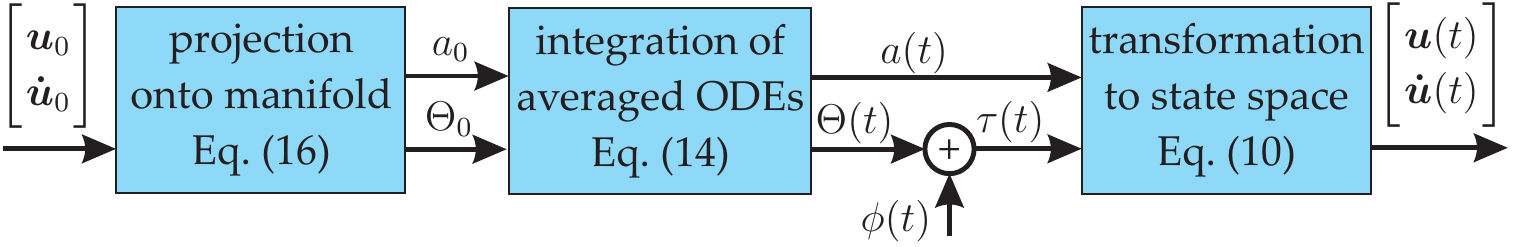}{Overview of the proposed algorithm for the
approximation of the slow dynamics of nonlinear modes}{.85}
In a typical technical problem, the initial values for \g{a,\Theta}
are not a priori known, but rather the initial values are given in
state space \g{\mms{u}_0,\mms{\dot u}_0}. In this paper, a closest
point projection is proposed to find a suitable point on the
manifold corresponding to the actual initial state. This projection
can be achieved by solving the following minimization problem,
\e{ a_0,\Theta_0 =
\underset{a,\Theta}{\operatorname{arg}\min}\begin{Vmatrix}\mms
u_0+\frac{1}{\ii\Omega}\mms{\dot u}_0-a\mms
v\left(a,\absphase\left(\Theta\right)\right)\end{Vmatrix}}{project}
It is generally possible that the initial state does not exactly lie
on the manifold. In this case, the energy is not confined to the
nonlinear mode anymore which violates the fundamental assumption of
the proposed ROM. Hence, agreement with the dynamic behavior of the
original system cannot be assured.\\
An overview of the proposed methodology is presented in
\fref{fig01}.

\section{Numerical examples\label{sec:numerical_examples}}
The authors developed a software environment for the nonlinear modal
analysis and ROMs for steady-state and transient predictions.
Several nonlinear example problems have been studied. The first
examples in \ssrefs{sdof}-\ssrefo{twodof} are single and two DOF
examples, respectively, and aim at demonstrating the capabilities of
the proposed method regarding complex problems in detail. The last
example in \ssref{beam} is a FE model of a beam with friction
contact and is designated to highlight the beneficial numerical
performance that can be achieved by using the ROM. The results of
the ROM are generally compared to the results obtained from direct
time integration of the original system.

\subsection{Singe degree-of-freedom systems\label{sec:sdof}}
\subsubsection{Duffing oscillator}
The dynamics of an autonomous, linearly damped Duffing oscillator
are investigated. The associated initial value problem reads
\e{\ddot u(t) + 2d\dot u(t) + u(t) + \knl u^3(t) = 0\fk\quad
u(0)=u_0,\,\dot u(0)=0\fp}{duffing}
For this problem, a closed-form analytical approximation can be
easily obtained by means of averaging \zo{nayf1979},
\e{u(t)\approx u_0\ee^{-dt}\cos\left(t+3\knl
u_0^2\frac{\ee^{-2dt}-1}{16}\right)\fp}{duffing_averaged_approx}
\myf[tbh]{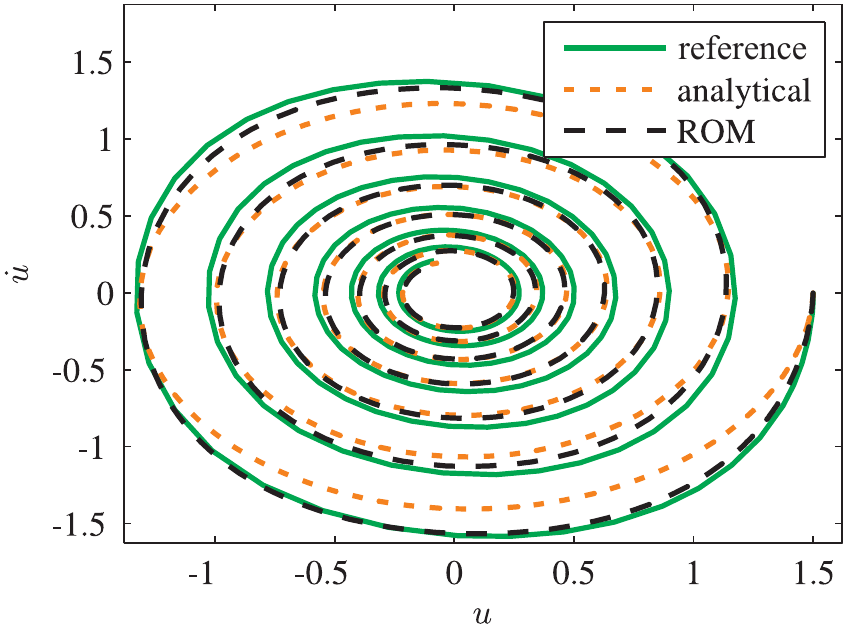}{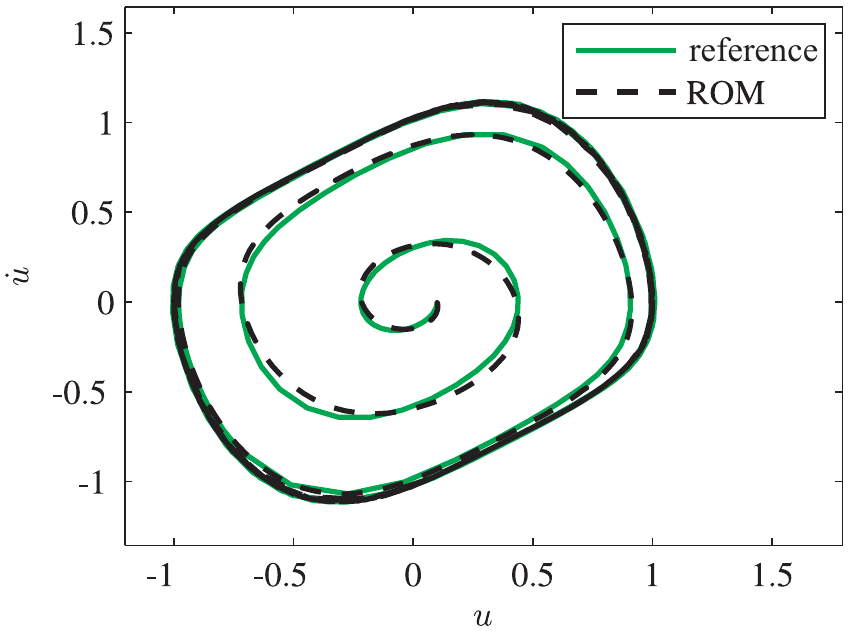}{}{}{.45}{.45}{Phase portraits of
autonomous single degree-of-freedom systems (~(a) Duffing oscillator
with linear damping, \g{d=0.05}, \g{\knl=0.25}, \g{u_0=1.5}, (b) Van
der Pol oscillator, \g{\alpha=0.5}, \g{\beta=2},\g{u_0=0.1}~)}
This analytical solution is compared to the proposed ROM and the
direct time integration results in the phase portrait in
\fref{fig02a}. It can be seen that the accuracy of the ROM is
significantly better that the analytical approximation for large
amplitudes. All three methods agree well for smaller amplitudes.
Slight deviations from the numerically computed reference solution
can be explained by the error introduced by averaging.

\subsubsection{Van der Pol oscillator}
Next, the autonomous Van der Pol oscillator is considered. The
initial value problem governing the autonomous dynamics can be
stated as
\e{\ddot u(t) - \left(\alpha-\beta u^2(t)\right)\dot u(t) + u(t) =
0\fk\quad u(0)=u_0,\,\dot u(0)=0\fp}{vanderpol}
In contrast to the previous example, the attractor is not a fixed
point but a periodic orbit known as limit cycle, see \fref{fig02b}.
Again, the results obtained by the ROM are in excellent agreement
with the reference method. Owing to the synthesis of all harmonics
contributing to the eigenvector in \eref{manifold}, the apparent
multiharmonic character of the limit cycle is well-captured.

\subsection{Two degree-of-freedom systems\label{sec:twodof}}
\fss[t!]{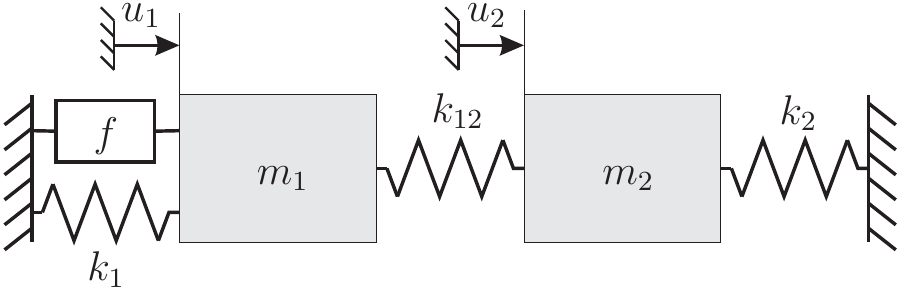}{Two degree-of-freedom system with nonlinear
element}{1.0}
In this subsection, a two DOF model is considered. It consists of
two masses and three linear springs, as illustrated in \fref{fig03}.
The effect of different nonlinear forces \g{f} acting on one of the
masses and stemming from either a cubic spring, a Coulomb friction
element or a unilateral spring, is investigated in the following. If
not otherwise specified, the results are illustrated for the
displacement \g{u_2}. Amplitude-dependent modal properties are
illustrated with respect to the modal amplitude \g{a_2} defined as
the fundamental harmonic amplitude of \g{u_2}, \ie
\g{a_2=\left|a\Psi_{1,2}\right|}. For the sake of clarity in the
figures, many results of the ROM are only depicted in terms of the
envelope.

\subsubsection{Cubic spring nonlinearity}
\fss[t!]{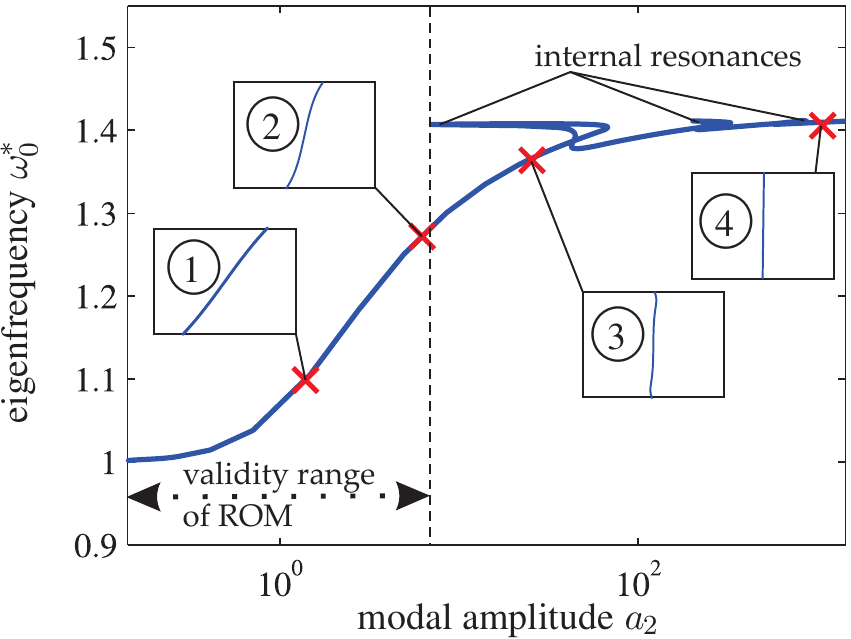}{Frequency-energy plot of the first nonlinear mode of
the system with cubic spring, subfigures represent phase projections
in the \g{u_2-u_1} plane, \g{m_1=m_2=1}, \g{k_1=k_2=k_{12}=1},
\g{\knl=0.5}}{1.0}
\fs[t!]{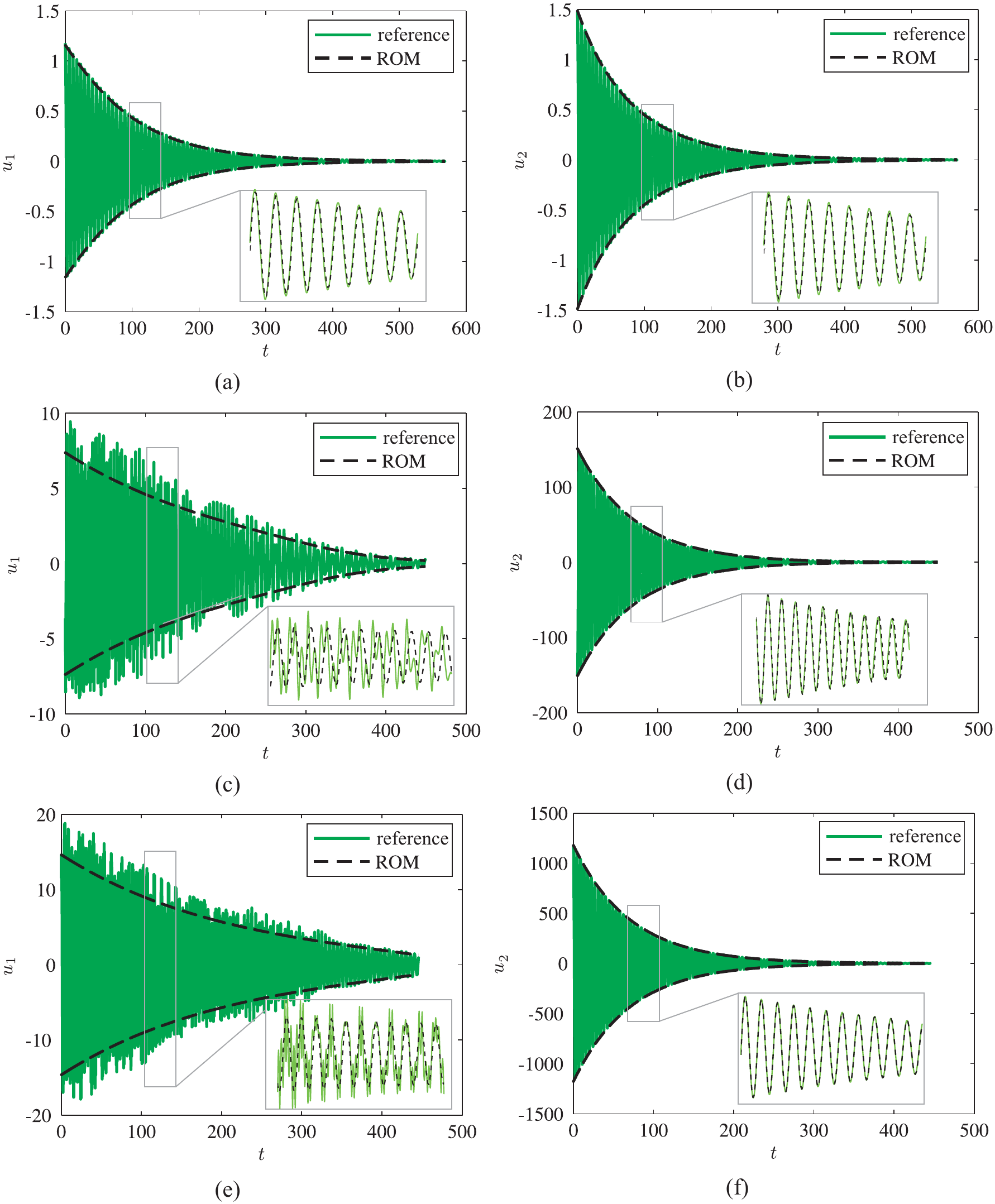}{Time histories of autonomous system with cubic
spring, \g{1\%} damping ratio (~(a) \g{u_1} for starting point
\kreis{1}, (b) \g{u_2} for starting point \kreis{1}, (c) \g{u_1} for
starting point \kreis{3}, (d) \g{u_2} for starting point \kreis{3},
(e) \g{u_1} for starting point \kreis{4}, (f) \g{u_2} for starting
point \kreis{4}~)}{1.0}
\fs[t!]{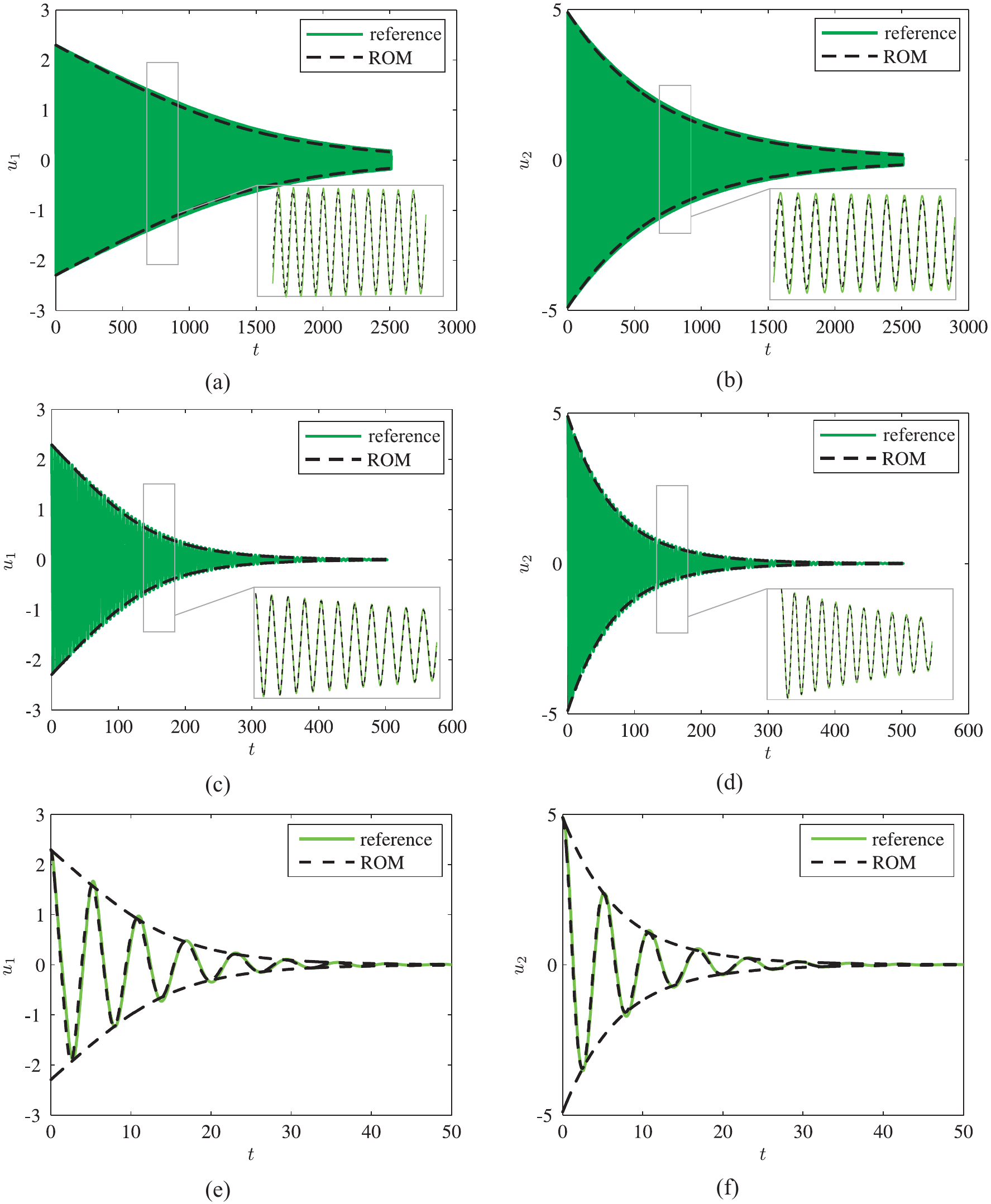}{Time histories of autonomous system with cubic
spring, starting point \kreis{2} (~(a) \g{u_1} for \g{0.1\%} damping
ratio, (b) \g{u_2} for \g{0.1\%} damping ratio, (c) \g{u_1} for
\g{1\%} damping ratio, (d) \g{u_2} for \g{1\%} damping ratio, (e)
\g{u_1} for \g{10\%} damping ratio, (f) \g{u_2} for \g{10\%} damping
ratio~)}{1.0}
In case of the nonlinear spring with stiffness \g{\knl}, the force
\g{f} reads
\e{f=\knl u_1^3\fp}{fnlcubic}
The so called Frequency-Energy-Plot is depicted for the first mode
in \fref{fig04}. Nonlinear modal interactions with the second mode
can be recognized in the form of tongues in \fref{fig04}
\zo{kers2009}. Not only the frequency increases with the modal
amplitude, but also the mode shape varies. For the points marked by
a cross, the mode shape is illustrated in \fref{fig04} in the
\g{u_2-u_1} plane. It can be seen that the mode shape becomes
nonlinear as the modal amplitude increases and finally localizes in
the right mass in \fref{fig03}. The interested reader is referred to
\zo{kers2009} for a
comprehensive stability and bifurcation analysis of this system.\\
We now focus on the transient, autonomous dynamics of this system.
To this end, a constant, mass-proportional damping \g{\mms{\tilde
C}\propto\mms M} is specified. The magnitude of the damping term
will be provided in terms of a linear (\ie for \g{f=0}) modal
damping ratio.\\
In \frefs{fig05}, the time histories of both masses are illustrated
for different initial modal amplitude values (and therefore also
mode shape) corresponding to the points indicated in \fref{fig04}.
The dynamics of the left mass (\g{u_1}) appear to be distorted
beyond the modal amplitude value where the first bifurcation of the
nonlinear mode occurs. This behavior cannot be predicted by the
single modal ROM followed in this study. As expected, the ROM is
restricted to modal amplitude regimes where the nonlinear modes do
not interact with each other \zo{kers2009,blan2013}. However, the
envelope of the displacement \g{u_2} of the right mass is in
excellent agreement with the reference simulation. This applies even
for large initial modal amplitudes, \ie regardless of the presence
of nonlinear modal
interactions.\\
In \fref{fig06}, the time histories are depicted for the same
initial modal amplitude (before the bifurcation point) but different
damping values. Again, good agreement between the ROM and the
reference results can be ascertained, in particular for displacement
\g{u_2}. The consideration of the effect of damping on the modal
properties in the ROM according to \eref{perturbedmodalproperties}
can therefore be regarded as valid for this example, even for
damping ratios as large as \g{10\%}.

\subsubsection{Coulomb friction nonlinearity}
The cubic spring is now replaced by a Coulomb friction element. The
nonlinear force characteristic is defined as follows,
\e{f = R\tanh\left(\frac{\dot x_1}{\epsilon}\right)\fp}{coulomb}
Herein, \g{R} is the limit friction force and \g{\epsilon} is a
small regularization parameter determining the accuracy of the
approximation of the signum function actually contained in the
Coulomb law \g{f=R~\sgn\dot x_1}. It should be pointed out that the
modal analysis results depicted with scaled \g{a/R} axis are
identical for any limit friction force value \g{R}
\zo{krac2013d,krac2013a}. This scaling property allows for a
straight-forward extension of the ROM parameter space at no extra
computational cost.
\myf[t!]{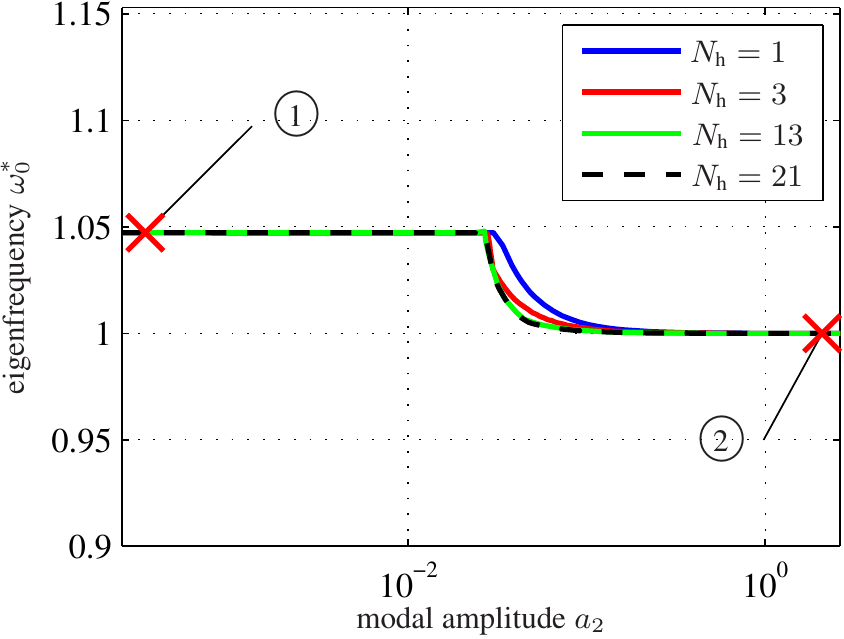}{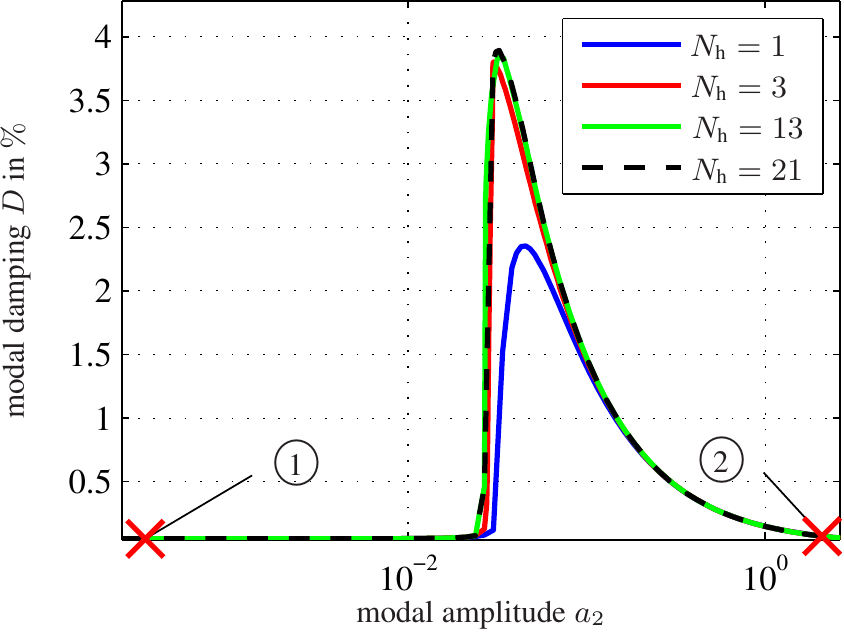}{}{}{.45}{.45}{Nonlinear modal properties of
first mode of system with friction nonlinearity, \g{m_1=0.02},
\g{m_2=1}, \g{k_1=0}, \g{k_{12}=40}, \g{k_2=600}, \g{R=1},
\g{\epsilon=0.01} (~(a) eigenfrequency, (b) modal damping~)}
\fss[t!]{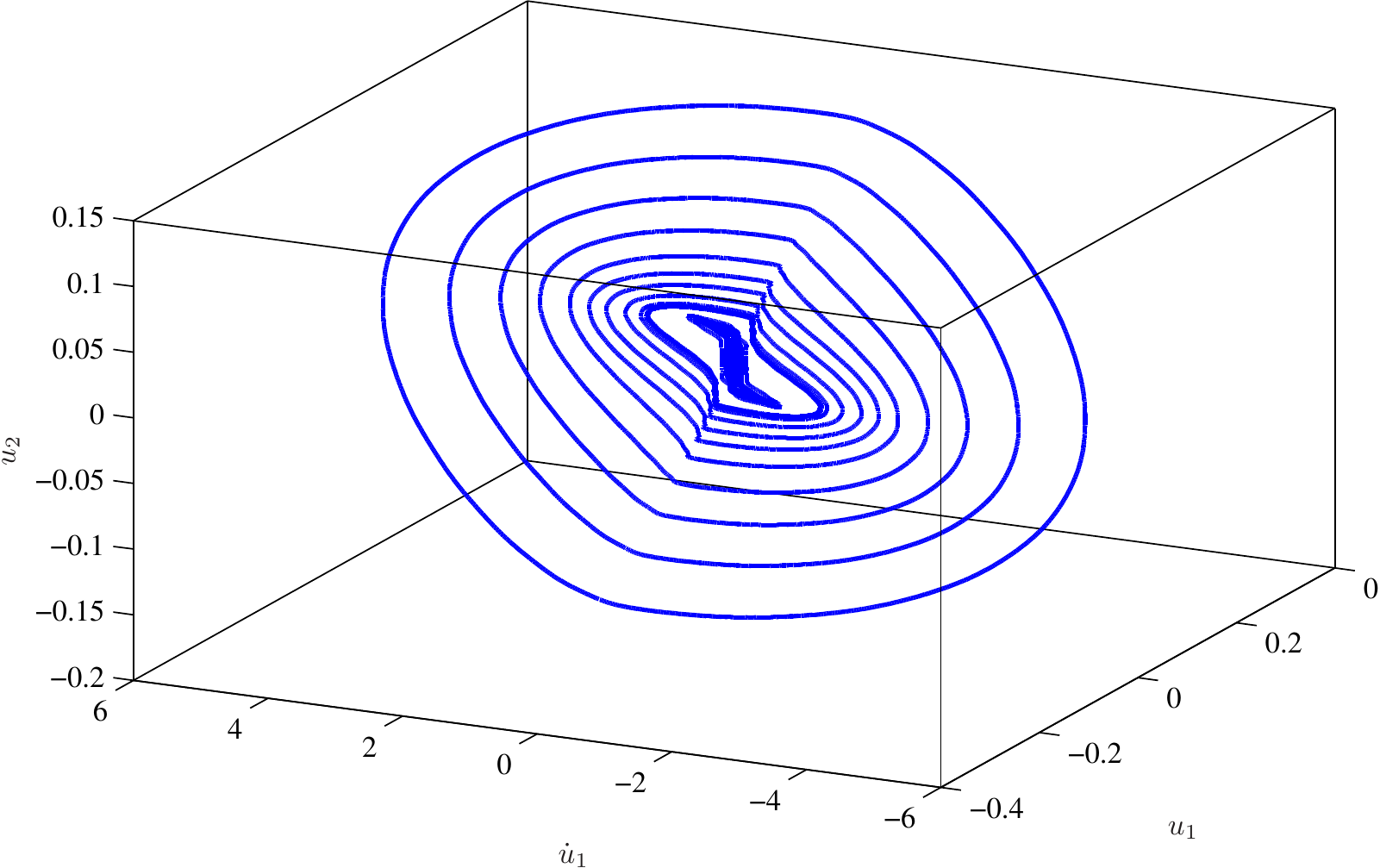}{Manifold of first mode of system with friction
nonlinearity}{.85}
\\
In \frefs{fig07a}-\frefo{fig07b}, eigenfrequency and modal damping
are illustrated for the first nonlinear mode of the system. The
system exhibits two linear limit cases: For low amplitudes, the
friction contact is fully stuck so that \g{u_1=0}. For very large
amplitudes, the friction contact is always sliding. The limited
friction force, however, has decreasing effect on the dynamic
properties for large amplitudes so that the modal properties
approach the values corresponding to the system without friction
element. Note that a moderate harmonic order \g{\nh} is required to
accurately capture the friction effect on this system. The
multiharmonic character of the eigenmode also becomes apparent in
the manifold plot in \fref{fig08}. The abrupt changes between stick
and slip motion make the mode shapes significantly deviating from
elliptic orbits. Furthermore, the mode shape significantly varies
with the modal amplitude.
\myf[t!]{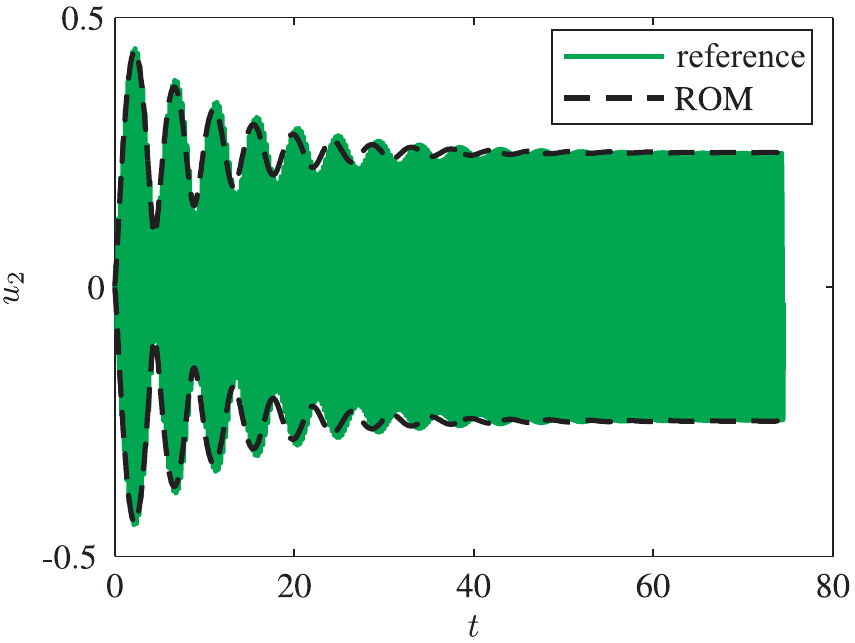}{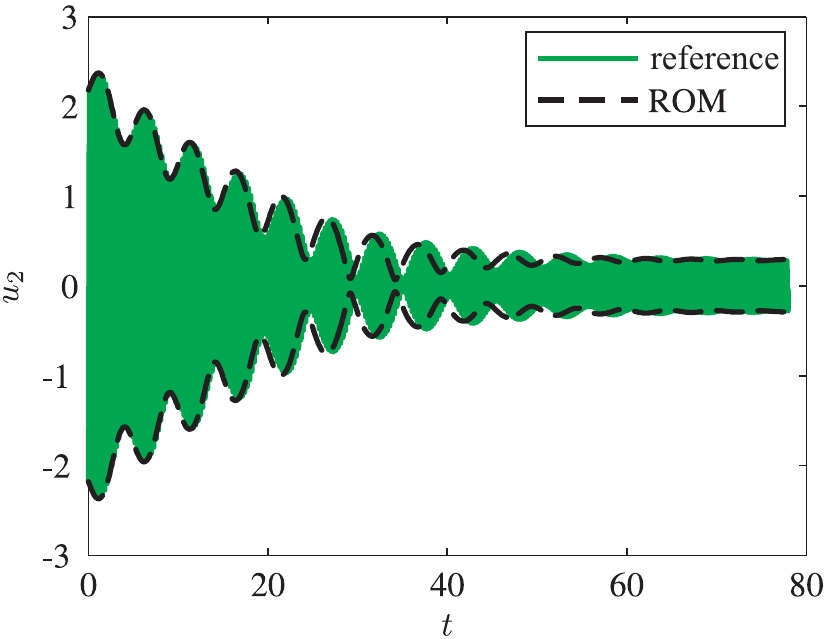}{}{}{.45}{.45}{Time histories of system with
friction nonlinearity subject to steady-state harmonic forcing (~(a)
frequency \g{\Omega=0.9\ommod(a=0)}, starting point \kreis{1}, (b)
frequency \g{\Omega=0.95\ommod(a=0)}, starting point \kreis{2}~)}
\fs[t!]{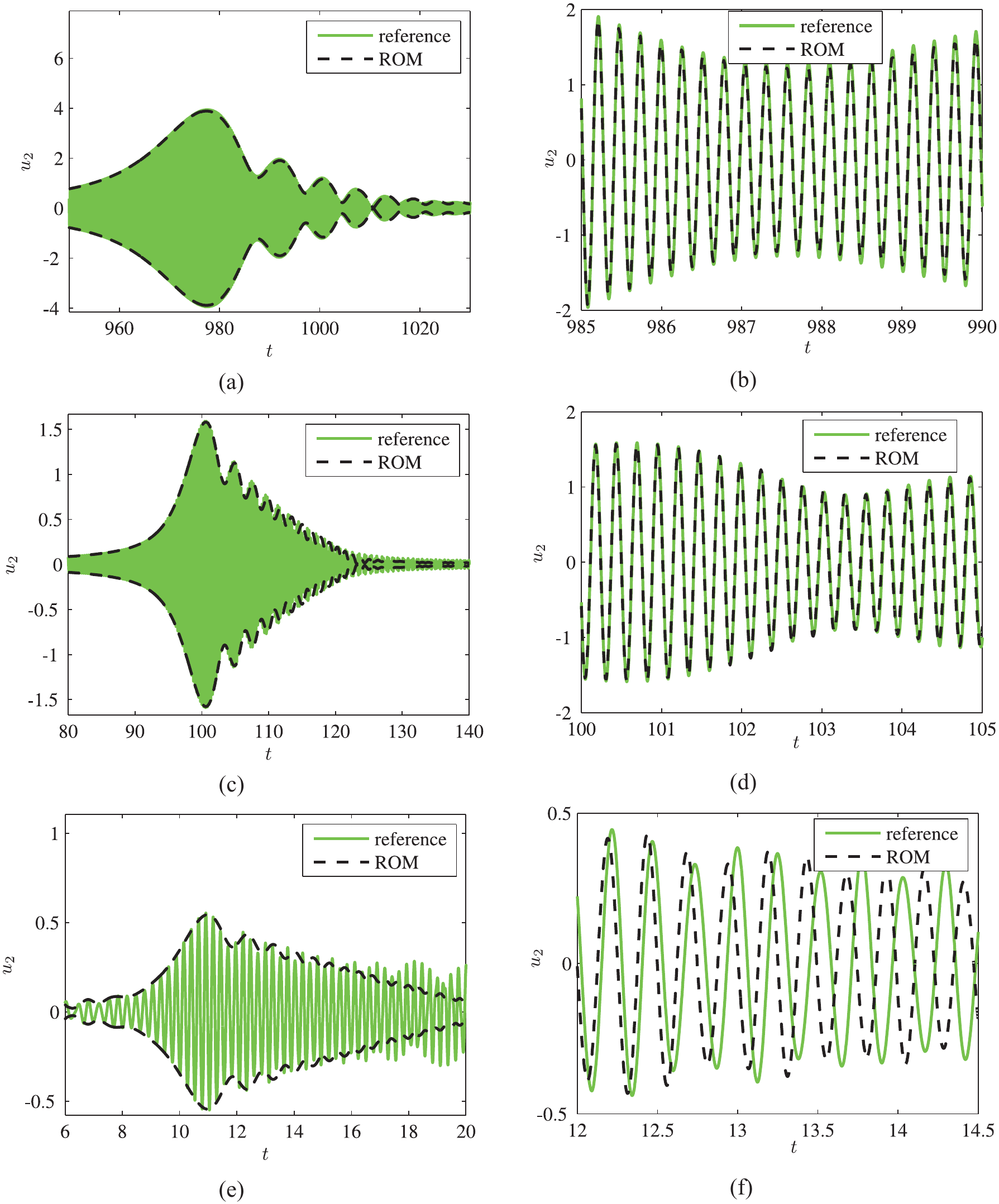}{Time histories of system with friction nonlinearity
subject to a sine sweep (~(a) and (b) \g{\dot\phi_{\mathrm
e}(t)=0.025~t}, (c) and (d) \g{\dot\phi_{\mathrm e}(t)=0.25~t}, (e)
and (f) \g{\dot\phi_{\mathrm e}(t)=2.5~t}~)}{1.0}
\\
The two-mass system with friction element is only investigated in
the forced configuration, for an autonomous system subject to
friction, see \ssref{beam}. In \frefs{fig09a}- \frefo{fig09b}, the
time histories are illustrated for the case of steady forcing with
near-resonant excitation frequency. Two different initial conditions
are considered: In \fref{fig09a}, the system starts from its
equilibrium point, while it starts from a modal amplitude larger
than the steady-state amplitude in
\fref{fig09b}.\\
Beating phenomena occur in both cases before reaching the
steady-state with a constant amplitude. According to expectations,
the frequency of this beating is smaller in \fref{fig09b} than it is
in \fref{fig09a} since eigenfrequency and excitation frequency are
closer in the first case. Dissipation due to friction and linear
damping is the reason for the decay of the pulsation of the envelope
part of the solution and the transition to the limit cycle. The
amplitude of the limit cycle depends on the excitation level.
Despite the initial energy of zero, it can be seen in \fref{fig09a}
that the amplitude overshoots the steady-state amplitude before
reaching the limit cycle. The slow flow results agree well with the
direct time integration results for both steady cases, in spite of
the strongly
nonlinear system behavior.\\
In \fref{fig10}, the time histories are depicted for the case of a
sine sweep, \ie quasi-harmonic forcing with linearly increasing
excitation frequency. In all three cases the system dynamics are
specified to
start from the equilibrium point.\\
Again, a pulsation phenomenon occurs in the response, which is
accurately captured by the proposed approximation method. The
maximum amplitude of the first pulse decreases with increasing
angular acceleration. The excitation frequency at this maximum
amplitude apparently increases with increasing angular acceleration.
In case of the largest frequency acceleration in \fref{fig10}(c),
the eigenfrequency of the second mode is reached within the depicted
time span. In full accordance with the restriction of the ROM, the
dynamics are only predicted accurately in the vicinity of the
eigenfrequency of the considered nonlinear mode, \ie the first mode
in this case. It is assumed that the accuracy of the proposed method
could be improved in this case by simply superimposing the ROMs for
both nonlinear modes. This has, however, not been done in the
present study. It should be noted that the maximum amplitude in the
near the first resonance, which is typically of importance for
design considerations, is accurately predicted by the ROM.

\subsubsection{Unilateral spring nonlinearity}
Next, the effect of a unilateral spring on the dynamics of the
system in \fref{fig03} is investigated. The unilateral spring of
stiffness \g{\knl} is considered to be preloaded by a constant force
\g{N} so that the nonlinear force reads
\e{f = \begin{cases} -N & \knl u_1<-N\\
\knl u_1 & \knl u_1\ge-N\end{cases}\fp}{fnlunilateral}
Eigenfrequency and mainfold of the second nonlinear mode are
depicted in \frefs{fig11a}-\frefo{fig11b}. Note that since the
system is conservative, the nonlinear modal damping is identical to
zero and therefore not depicted. As soon as the amplitudes are large
enough, the preload is exceeded so that the spring undergoes
lift-off during the period of oscillation. As a consequence, the
nonlinear mode becomes asymmetrical to the origin. This can be
easily deduced from \fref{fig11b}. It is thus essential to not only
account for the higher harmonics but also to consider the zeroth
harmonic of the nonlinear mode.
\myf[t!]{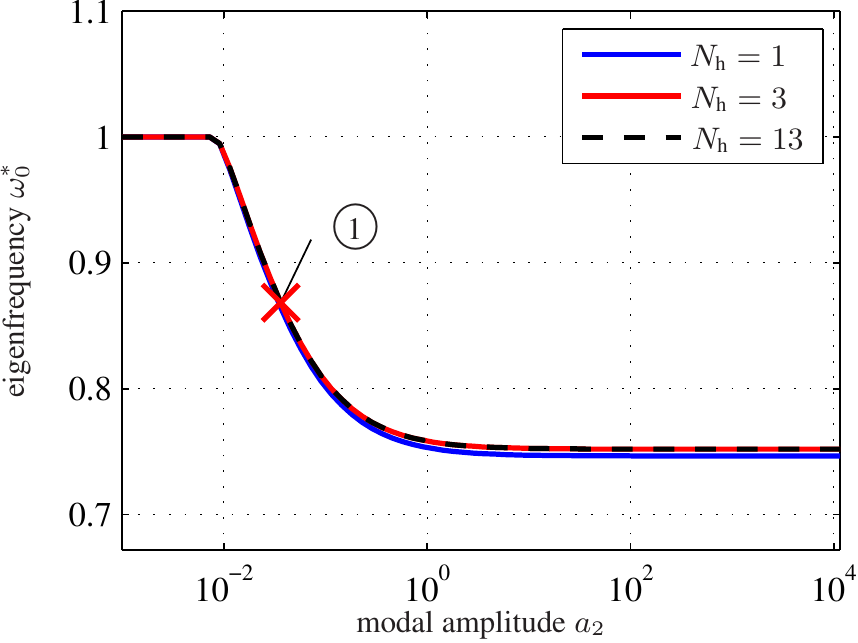}{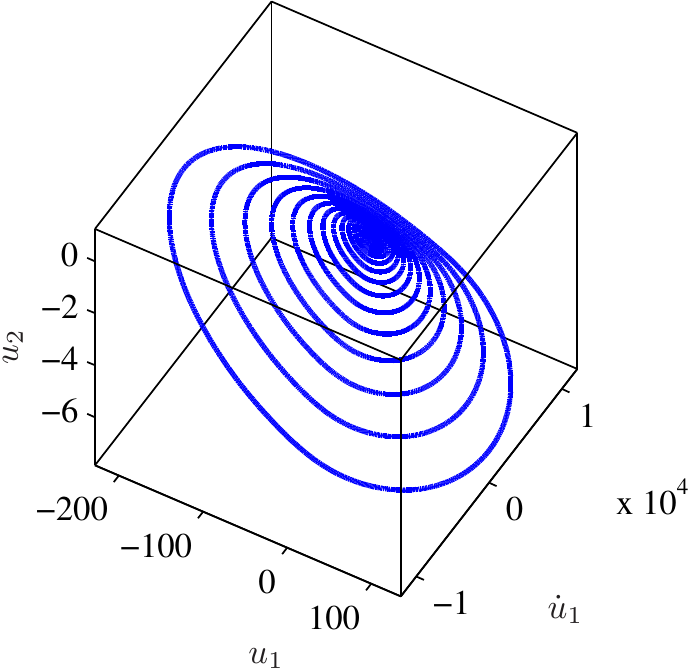}{}{}{.45}{.45}{Nonlinear modal properties of
second mode of system with unilateral spring, \g{m_1=0.02},
\g{m_2=1}, \g{k_1=0}, \g{k_{12}=40}, \g{k_2=600}, \g{N=1/70},
\g{\knl=70} (~(a) eigenfrequency, (b) manifold~)}
\\
The system is first investigated in the autonomous configuration.
Starting from a moderate initial amplitude as indicated in
\fref{fig11a}, the system approaches its fix point as illustrated in
the time histories in \fref{fig12}. A constant, mass-proportional
damping has been specified such that all modes have a damping ratio
of \g{1\%} in the linearized case. The effect of the stiffness term
\g{\mms{\tilde K}} in the ROM is investigated.
\fs[t!]{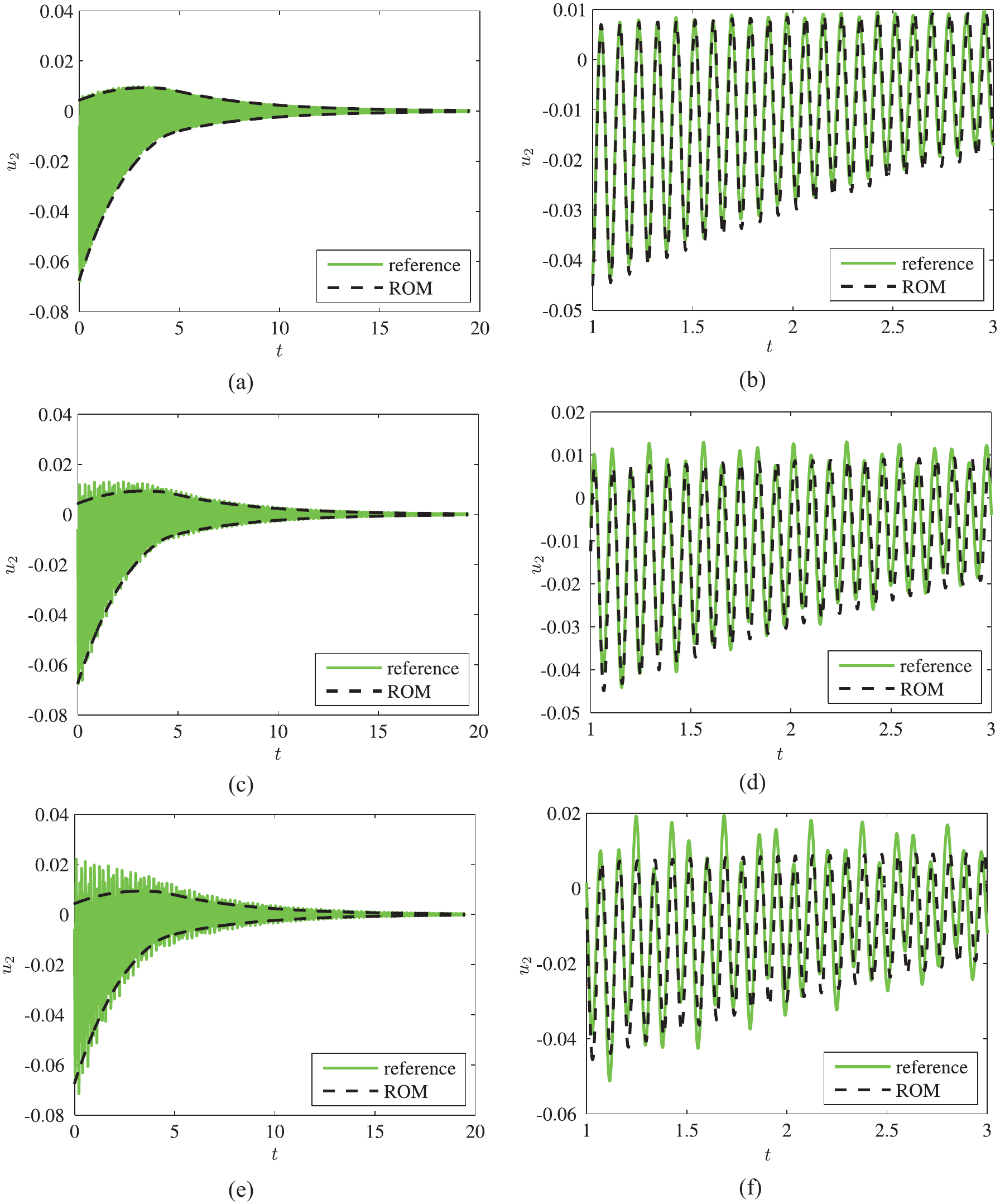}{Time history of autonomous system with unilateral
spring, starting point \kreis{1} (~(a) and (b) \g{\mms{\tilde
K}=\mms 0}, (c) and (d) \g{\mms{\tilde K}=0.1~\mms K}, (e) and (f)
\g{\mms{\tilde K}=0.25~\mms K}~)}{1.0}
\myf[t!]{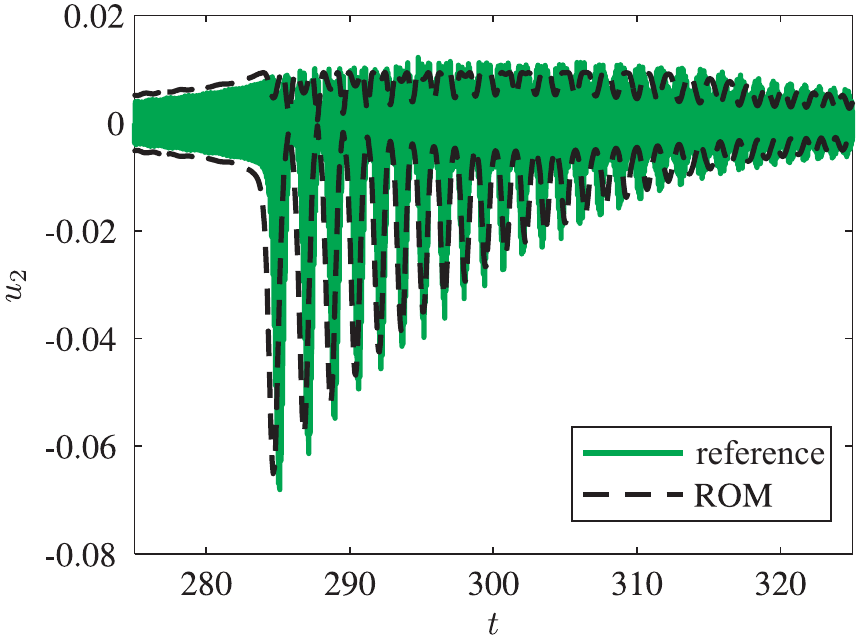}{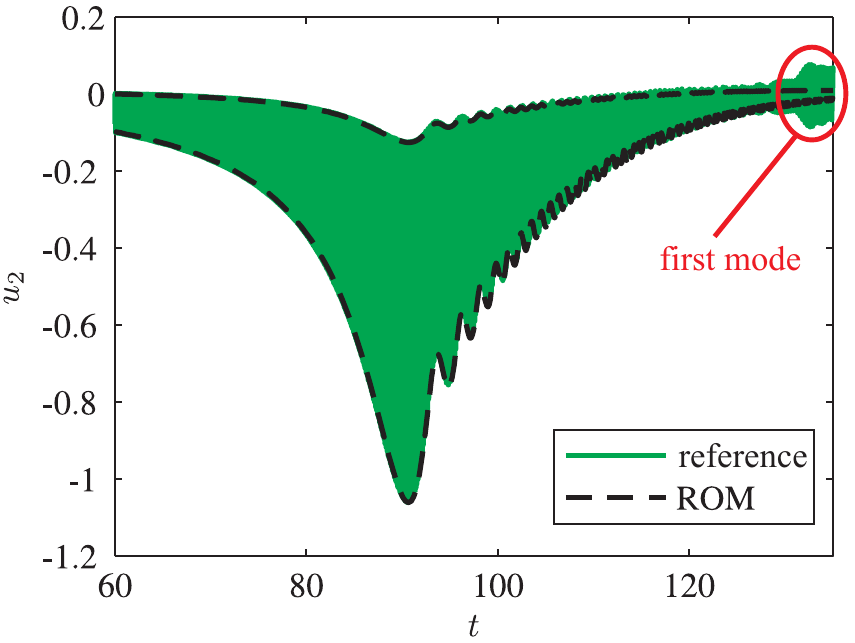}{}{}{.45}{.45}{Time history of system with
unilateral spring subject to a sine sweep (~(a) \g{\dot\phi_{\mathrm
e}(t)=0.25~t}, (b) \g{\dot\phi_{\mathrm
e}(t)=1.75\ommod(a=0)-0.025~t}~)}
\\
In the case \g{\mms{\tilde K}=\mms 0}, the proposed method is in
excellent agreement with the direct time integration, see upper two
images in \fref{fig12}. Note the asymmetrical upper and lower
envelope of the response. In accordance with the manifold in
\fref{fig11b}, there is an offset such that the mean value of
\g{u_2} is negative in the region of partial lift-off. As the
amplitude decreases due to the linear damping, the relative mean
value decreases and finally vanishes as the lift-off phases become
shorter within one cycle of oscillation and finally vanish.\\
In case of nonzero stiffness terms \g{\mms{\tilde K}\neq\mms 0}, the
response appears to be distorted, see lower four images in
\fref{fig12}. It is assumed that this distortion mainly results from
the fact that the initial state is not precisely on the invariant
manifold of the perturbed system. The averaged results can therefore
only approximate the mean
envelope in this case.\\
In \frefs{fig13a}- \frefo{fig13b}, the transient dynamics of the
system subject to a sine sweep is illustrated. Once more, a very
good agreement between proposed approximation and the direct time
integration can be stated in the
vicinity of the first eigenfrequency.\\
The asymmetrical character with respect to amplitude is again
induced by the unilateral nonlinearity can easily be seen from the
results. Similar to the results in \fref{fig10}(c), the ROM fails in
predicting the `blast' occurring at large times in \fref{fig13b}.
Here, the excitation frequency reaches the eigenfrequency of the
first mode an drives the mode into resonance. Since only the second
mode was
considered in the ROM, this phenomenon is not predicted.\\
It is noteworthy that there exists a strong qualitative discrepancy
between \fref{fig13a} and \fref{fig13b}. This discrepancy is not
only caused by the deviation in the magnitude but also the sign of
the angular acceleration. A typical frequency response curve of a
system with unilateral preloaded spring is bent to the left such
that there exists a frequency range with a multi-valued response.
Hence, the results for a down-sweep generally deviate from those of
an up-sweep. Starting from a subcritical (supercritical) frequency,
a jump phenomenon occurs when the frequency is increased (decreased)
beyond the folding point of the frequency response curve.
Comparatively fast amplitude changes can also be observed from
\frefs{fig13a}- \frefo{fig13b}. However, the jump is not severe in
this case due to the finiteness of the angular acceleration.

\subsection{Beam with friction nonlinearity\label{sec:beam}}
\fss[b!]{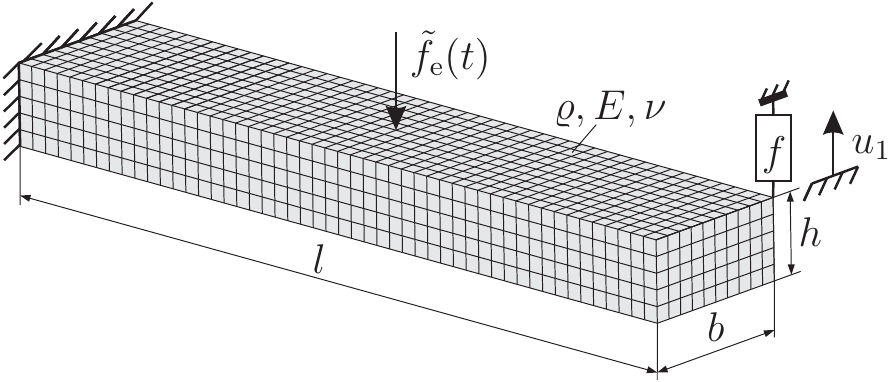}{Cantilevered beam with nonlinear element subject to
excitation}{1.0}
\myf[htb]{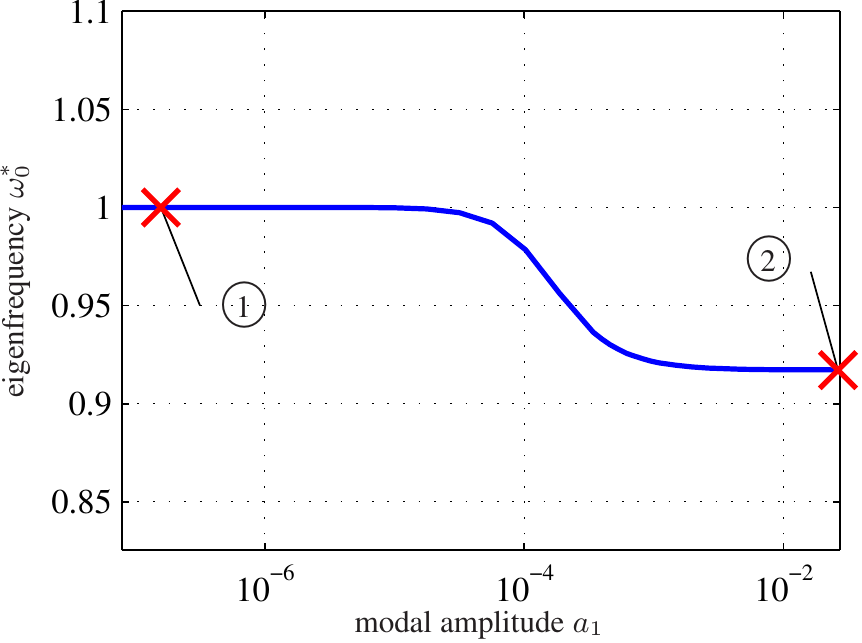}{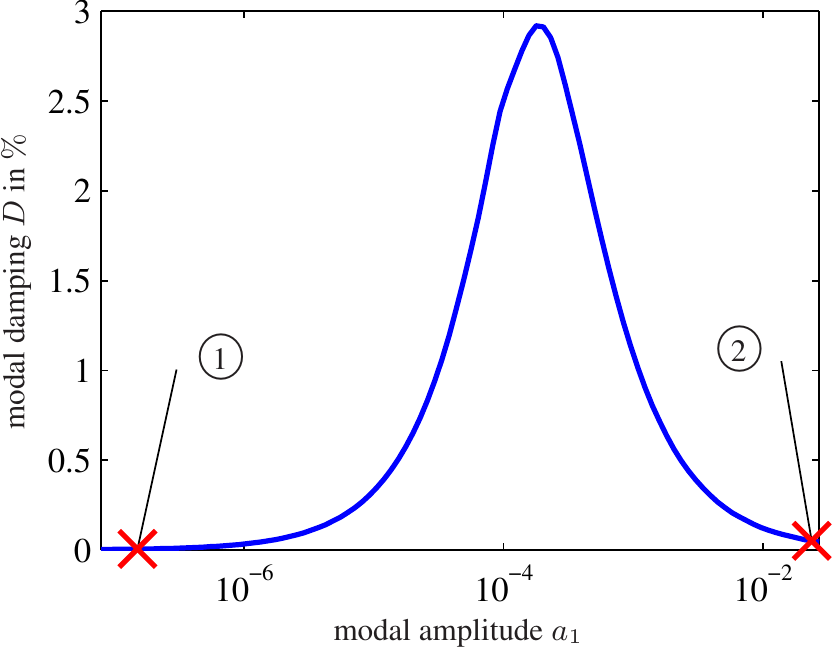}{}{}{.45}{.45}{Nonlinear modal properties
of first bending mode of beam with friction nonlinearity,
\gehg{\rho}{4430}{kg/m^3}, \gehg{E}{100}{GPa}, \g{\nu=0.3},
\gehg{l}{1}{m}, \gehg{b}{0.2}{m}, \gehg{h}{0.1}{m},
\gehg{\kt}{1}{kN/mm}, \gehg{R}{100}{N}, \g{\alpha=1} (~(a)
eigenfrequency, (b) modal damping~)}
%
%
\myf[htb]{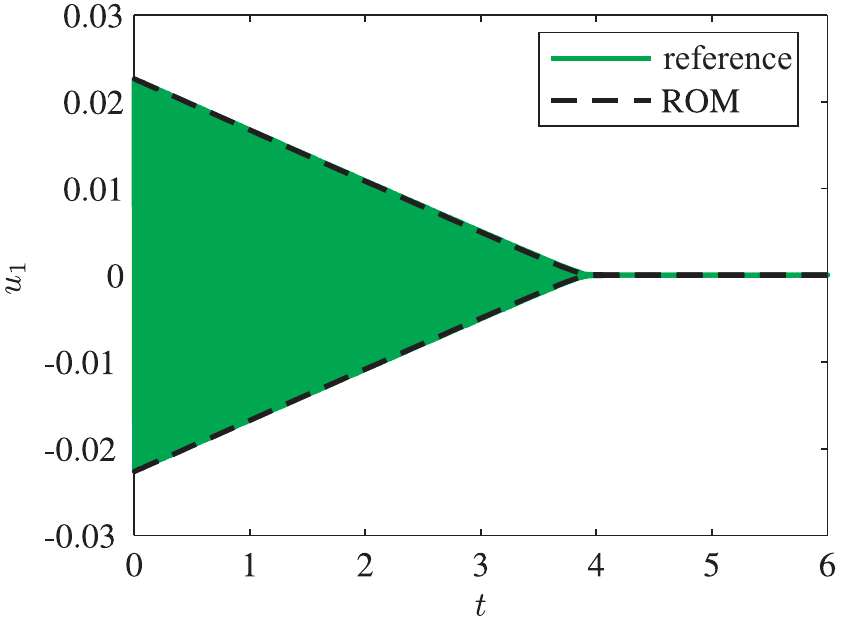}{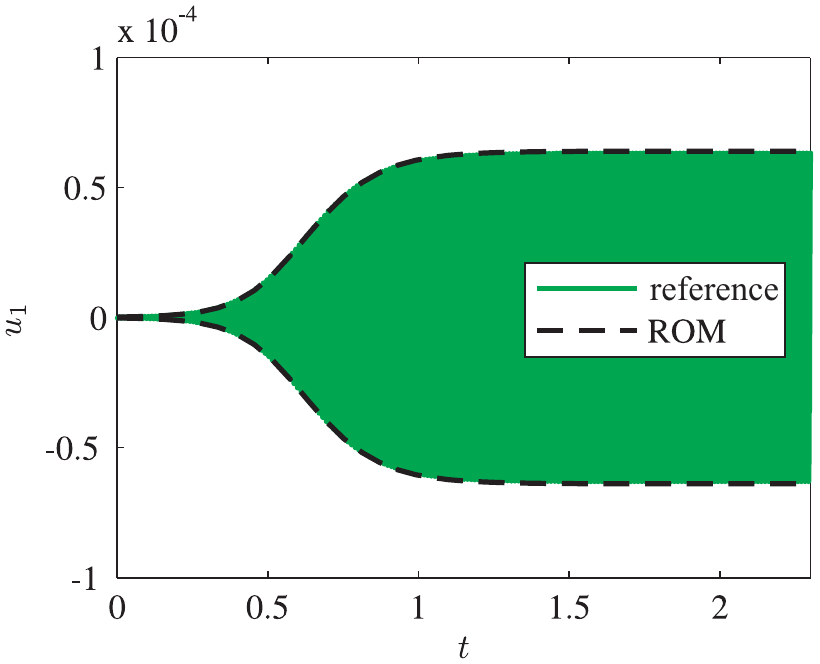}{}{}{.45}{.45}{Time history of autonomous
beam with friction nonlinearity (~(a) transition to equilibrium with
\g{\mms{\tilde C}=\mms 0} from starting point \kreis{1}, (b)
transition to limit cycle with \g{\mms{\tilde C=\mms{\tilde
C}_{-2\%}}} from starting point \kreis{2}~)}
\myf[htb]{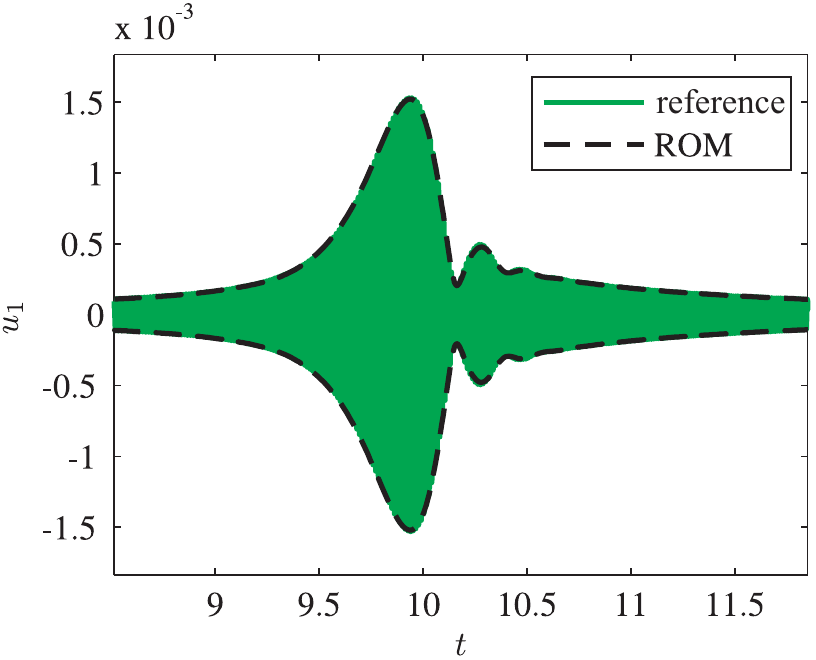}{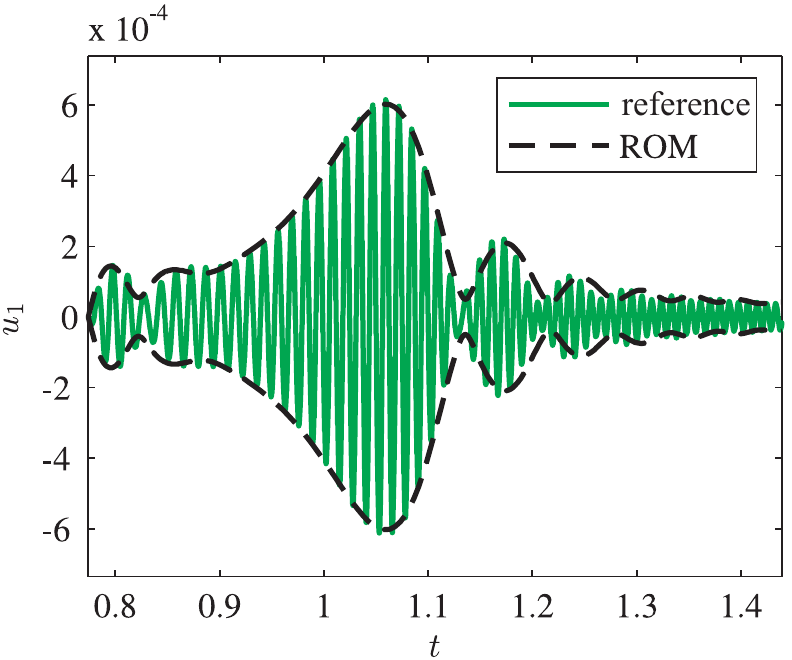}{}{}{.45}{.45}{Time history of beam with
friction nonlinearity subject to a sine sweep (~(a) slow run-up,
\g{\dot\phi_{\mathrm e}(t)=50~t}, (b) fast run-up,
\g{\dot\phi_{\mathrm e}(t)=500~t}~)}
A clamped beam with friction nonlinearity at single node at its free
end has been investigated, see \fref{fig14}. The geometry was
spatially discretized by means of solid Finite Elements. The
original model comprised \g{3366} nodes and \g{9900} DOFs. The
Craig-Bampton technique was used to reduce the order of the
underlying linear model. The first bending mode was studied. In
accordance with a preliminary convergence study, the system dynamics
are accurately described in the reduced basis composed of the static
constraint mode of the
nonlinear DOF and the first five fixed interface normal modes.\\
The Dahl friction model \zo{dahl1976} is considered. The nonlinear
friction force is governed by a differential equation,
\e{\dot f = \kt\left(1-\frac{f}{R}\sgn\dot u_1\right)^\alpha\dot
u_1\fp}{dahl}
Herein, \g{\kt} is the initial slope of the hysteresis, \g{R} is the
limit friction force and \g{\alpha} is a parameter determining the
shape of the hysteresis.\\
In \frefs{fig15a}-\frefo{fig15b}, eigenfrequency and modal damping
are depicted for the first mode of the system. The modal amplitude
\g{a_1} defined as the fundamental harmonic amplitude of the
displacement \g{u_1} of the contact node, \ie
\g{a_1=\left|a\Psi_{1,1}\right|}. The results are generally similar
to the ones obtained for the two DOF system with Coulomb
nonlinearity, see \frefs{fig07a}-\frefo{fig07b}. Since the Dahl
model also describes the microslip behavior, the modal properties
are
smoother compared to results for the Coulomb macroslip model.\\
The system is first considered in the autonomous configuration
without external forcing. In the time history in \fref{fig16a}, the
transient dynamics towards the equilibrium point are depicted. The
well-known linear amplitude
decay can be clearly deduced from the results.\\
The effect of aerodynamic instabilities such as flutter on the
vibration behavior of nonlinear mechanical structures can be
approximately described by an indefinite linear damping matrix
\zo{petr2012c,krac2013d}. This can result in so called
flutter-induced limit cycle oscillations. In this case, a constant
damping matrix \g{\mms{\tilde C}_{-2\%}} has been specified in such
a way that the linearized system has a negative damping ratio of
\g{-2\%} for the first mode and a positive damping ratio of \g{1\%}
for the remaining modes. The transition from zero amplitude to a
stable limit cycle is depicted in \fref{fig16b}. Again, the results
of the proposed ROM for the autonomous system are in
excellent agreement with the direct time integration results.\\
The system is next considered in the heteronomous configuration with
a sine sweep excitation. The time histories for two different
frequency acceleration values are depicted in \frefs{fig17a}-
\frefo{fig17b}. The results are generally in very good agreement.
According to expectations, the agreement is slightly worse for
considerably large frequency acceleration values, \ie when the fast
and slow phase have similar time scales. However, the time and
amplitude of the largest peak as well as the qualitative modulation
behavior is captured well by the proposed ROM.
\tab[htb]{r|cc|cc}{Problem & ODE dim. & CPU time & ODE dim. & CPU time\\
& (direct) & (direct) & (ROM) & (ROM)\\
\hline\hline autonomous (\fref{fig16a}) & $13$ & $89~\rm s$ & $1$ &
$0.1~\rm s$\\
autonomous (\fref{fig16b}) & $13$ & $13~\rm s$ & $1$ &
$0.1~\rm s$\\
forced (\fref{fig17a}) & $13$ & $104~\rm s$ & $2$ &
$2.2~\rm s$\\
forced (\fref{fig17b}) & $13$ & $14~\rm s$ & $2$ & $0.7~\rm
s$}{Computational effort for conventional and proposed
methodology}{comp_effort}
\\
The computational effort required by the ROM is compared to the
direct time integration of the original system in
\tref{comp_effort}. The problem dimension of the original system is
two times the number of retained generalized DOFs plus one dimension
for the differential equation governing the friction effect in
\eref{dahl}, \ie the system dimension is \g{2\cdot 6+1=13}. In case
of the ROM the problem dimension is two (amplitude \g{a} and slow
phase \g{\Theta}) in the heteronomous case and one (amplitude \g{a})
in the autonomous case.\\
It can be easily ascertained from \tref{comp_effort} that the
computational effort to obtain the time histories presented in this
subsection can significantly be reduced by the proposed ROM. Not
only the decreased problem dimension contributes to a reduction of
the computational effort but also the averaging approach itself
facilitates larger time steps in the integration process: The
governing differential equations in \eref{odeslowflow} are
formulated in the apparently more slowly varying coordinates and
also the involved nonlinear terms are typically smoother compared to
the ones involved in the original problem.


\section{Conclusions\label{sec:conclusions}}
A novel method for the numerical computation of the slow dynamics of
nonlinear mechanical systems has been developed. The method consists
of a two-step procedure: In the first step, a multiharmonic analysis
of the autonomous system is performed to directly compute the
amplitude-dependent characteristics of the considered nonlinear
mode. In the second step, these modal properties are used to
construct a two-dimensional reduced order model (ROM). The numerical
examples showed that the proposed ROM can be applied to various
problems including strongly nonlinear conservative as well as
non-conservative mechanical systems. The ROM is capable of directly
calculating steady-state and approximating the slow dynamics of
these systems in autonomous and heteronomous configurations provided
that the vibration energy is confined to an isolated nonlinear mode.
Moreover the ROM features a large parameter space including
additional linear damping, stiffness and near-resonant forcing
terms.\\
The proposed concept reduces the gap between often studied academic
single-degree-of-freedom and industrial FE models for nonlinear
dynamical problems. It is believed that the ROM developed in this
study represents a good basis for many multiphysics and
multi-component problems, where the ROM is used to describe the
structural dynamics of specific components. Further, the successful
prediction of unsteady dynamics such as the energy decay of
nonlinear modes is regarded as a corner stone for experimental
nonlinear modal analysis of dissipative systems.\\
Future work should include the extension of the proposed ROM to
problems where modal interactions of a finite number of nonlinear
modes occur. Moreover, an error estimation of the ROM compared to
the original model is highly desirable. In particular, the error
induced by damping and stiffness terms that are only considered in
the ROM, or starting points that lie outside the invariant manifold
should be investigated in more detail.

\appendix
\section{Derivation of the averaged equations governing the slow dynamics of a nonlinear mode\label{asec:appendix}}
In order to apply the complexification-averaging technique to
\eref{eqm_perturbed}, the acceleration \g{\mms{\ddot u}(t)} has to
be expressed in terms of the complex function \g{a\mms v}. This can
be achieved by noting that \g{a\mms v = \mms u + \frac{\mms{\dot
u}}{\ii\Omega}} from \eref{complexification} and taking the
derivative with respect to time:
\ea{a\mms v &=& \mms u + \frac{\mms{\dot u}}{\ii\Omega}\\
\Rightarrow \dot{\left(a\mms{v}\right)} &=& \mms{\dot u} +
\frac{\mms{\ddot
u}}{\ii\Omega}\\
\Rightarrow \mms{\ddot u} &=&
\ii\Omega\left(\dot{\left(a\mms{v}\right)}-\mms{\dot
u}\right)\fp}{uddot}
The time derivative \g{\dot{\left(a\mms{v}\right)}} is computed from
\erefs{manifold}{phase}, taking into account the variation of the
slow phase \g{\dot\Theta} and amplitude \g{\dot a},
\e{\dot{\left(a\mms{v}\right)} = \suml{n=0}{\nh}\mms\Psi_n\ee^{\ii
n\absphase}\left(\dot a + \ii n
\left(\Omega+\dot\Theta\right)a\right) +
\underbrace{\frac{\partial\mms\Psi_n}{\partial a}\dot a a \ee^{\ii
n\absphase}}_{\approx 0}\fp}{vdot}
The term associated with the sensitivity of the eigenvector and the
slow variation of the amplitude is considered a second-order effect
and therefore neglected in this study. By substituting \eref{vdot}
and the second equation in \eref{complexification} into
\eref{uddot}, one obtains the final form of \eref{uddotend}:
\e{\mms{\ddot u} = a\Omega^2\frac{\mms{v} - \mms{\overline v}}{2} +
\suml{n=0}{\nh}\mms\Psi_n\ee^{\ii n\absphase}\left(\ii\Omega\dot a -
n\Omega\left(\Omega+\dot\Theta\right)a\right)\fp}{uddotend}
\erefs{uddotend}{complexification} are then substituted into the
original equation of motion in \eref{eqm_perturbed},
\e{\mms{M}\ddot{\mms{u}}+ \mms{f}
=\mms\varepsilon\fk}{eqm_perturbed_appendix}
with the function \g{\mms\varepsilon} defined as,
\e{\mms\varepsilon(\mms u, \mms{\dot u}, t) = -\mms{\tilde K}\mms u
- \mms {\tilde C}\mms{\dot u} + \fehat~\frac{\ee^{\ii\phi_{\mathrm
e}(t)}+\ee^{-\ii\phi_{\mathrm e}(t)}}{2}\fp}{perturbationfunction}
Functional dependencies are dropped for the sake of brevity in
\eref{eqm_perturbed_appendix}. The resulting equation is then
projected onto the fundamental harmonic of the nonlinear mode
\g{\mms\Psi_1\ee^{\ii\absphase}}. Therefore, the inner product
\g{\mms{\Psi}_1\herm\langle\cdot,\ee^{\ii\absphase}\rangle} as
defined in \eref{inner_product} is applied to
\eref{eqm_perturbed_appendix}. The individual terms obtained by this
projection read
\ea{\mms{\Psi}_1\herm\langle\mms M\mms{\ddot
u},\ee^{\ii\absphase}\rangle &=& \ii\Omega\dot a -
\frac{\Omega^2}{2}a-\dot\Theta\Omega a\fk\\
\mms{\Psi}_1\herm\langle\mms f,\ee^{\ii\absphase}\rangle &=&
\frac{\ommod^2}{2}a+\dmod\ommod \ii\Omega a\fk\\
\mms{\Psi}_1\herm\langle\mms \varepsilon,\ee^{\ii\absphase}\rangle
&=& - \frac{\mms{\Psi}_1\herm\mms{\tilde K}\mms\Psi_1}{2}a -
\frac{\mms{\Psi}_1\herm\mms{\tilde C}\mms{\Psi}_1}{2}\ii\Omega a +
\frac{\mms{\Psi}_1\herm\mms{\fehat}}{2}\ee^{-\ii\Theta}\fp}{muddproj}
Herein, the normalization constraint in \eref{complex_evp},
\g{\mms\Psi_1\herm\mms M\mms\Psi_1=1}, was taken into account. The
projection of the nonlinear force \g{\mms f} is expressed in terms
of the modal properties of the corresponding nonlinear mode in full
accordance with \eref{complex_evp}. This approximation of the
nonlinear forces is the key aspect for the efficient ROM since the
often expensive nonlinear operator does not have to be evaluated in
ROM, but the readily available modal
properties are considered instead.\\
By substituting the projected terms in \eref{muddproj} into
\eref{eqm_perturbed_appendix} and equating real and imaginary parts,
one finally arrives at the ODE system in
\erefs{odeslowflow}{perturbedmodalproperties}.\\
It is interesting to note that only the complex eigenfrequency and
the fundamental harmonic \g{\mms\Psi_1} of the eigenvector occurs in
\eref{odeslowflow}. Due to the nonlinear character of
\eref{complex_evp}, these results are coupled to the remaining
harmonic components \g{\mms\Psi_n} and generally differ from the
results of a single-harmonic analysis.


\begin{thebibliography}{10}
\expandafter\ifx\csname url\endcsname\relax
  \def\url#1{\texttt{#1}}\fi
\expandafter\ifx\csname urlprefix\endcsname\relax\def\urlprefix{URL
}\fi \expandafter\ifx\csname href\endcsname\relax
  \def\href#1#2{#2} \def\path#1{#1}\fi

\bibitem{shaw1993}
S.~W. Shaw, C.~Pierre, {N}ormal {M}odes for {N}on-{L}inear
{V}ibratory
  {S}ystems, Journal of Sound and Vibration 164~(1) (1993) 85--124.

\bibitem{nayf2000}
A.~H. Nayfeh, {N}onlinear {I}nteractions: {A}nalytical,
{C}omputational and
  {E}xperimental {M}ethods, John Wiley {\&} Sons, 2000.

\bibitem{jian2005}
D.~Jiang, C.~Pierre, S.~W. Shaw, {N}onlinear normal modes for
vibratory systems
  under harmonic excitation, Journal of Sound and Vibration 288~(4-5) (2005)
  791--812.

\bibitem{pierre2006}
C.~Pierre, D.~Jiang, S.~W. Shaw, {N}onlinear normal modes and their
application
  in structural dynamics, Mathematical Problems in Engineering 10847~(2006), 1--15.

\bibitem{touz2006}
C.~Touz{\'e}, M.~Amabili, {N}onlinear normal modes for damped
geometrically
  nonlinear systems: {A}pplication to reduced-order modelling of harmonically
  forced structures, Journal of Sound and Vibration 298~(4--5) (2006) 958--981.

\bibitem{jian2004}
D.~Jiang, C.~Pierre, S.~W. Shaw, {L}arge-amplitude non-linear normal
modesof
  piecewise linear systems, Journal of Sound and Vibration 272~(3-5) (2004)
  869--891.

\bibitem{rens2013}
L.~Renson, G.~Kerschen, {N}onlinear {N}ormal {M}odes of
{N}onconservative
  {S}ystems, Proceedings of IMAC 31th Society of
  Experimental Mechanics Inc, February 11-14, Garden Grove, CA, USA (2013), 1--16.

\bibitem{chon2000a}
Y.~H. Chong, M.~Imregun, {D}evelopment and {A}pplication of a
{N}onlinear
  {M}odal {A}nalysis {T}echnique for {M}{D}{O}{F} {S}ystems, Journal of
  Vibration and Control 7~(2) (2000) 167--179.

\bibitem{gibe2003}
C.~Gibert, {F}itting measured frequency response using non-linear
modes,
  Mechanical Systems and Signal Processing 17~(1) (2003) 211--218.

\bibitem{kers2005}
G.~Kerschen, J.-c. Golinval, A.~F. Vakakis, L.~Bergman, {T}he
{M}ethod of
  {P}roper {O}rthogonal {D}ecomposition for {D}ynamical {C}haracterization and
  {O}rder {R}eduction of {M}echanical {S}ystems: {A}n {O}verview, Nonlinear
  Dynamics 41~(1) (2005) 147--169.

\bibitem{lee2010}
Y.~S. Lee, A.~F. Vakakis, D.~M. McFarland, L.~A. Bergman, {A}
global--local
  approach to nonlinear system identification: {A} review, Structural Control
  and Health Monitoring 17~(7) (2010) 742--760.

\bibitem{leun1992}
A.~Y. Leung, {N}onlinear modal analysis of frames by the incremental
  harmonic-balance method, Dynamics and Stability of Systems 7~(1) (1992)
  43--58.

\bibitem{ribe2000}
P.~Ribeiro, M.~Petyt, {N}on-linear free vibration of isotropic
plates with
  internal resonance, International Journal of Non-Linear Mechanics 35~(2)
  (2000) 263--278.

\bibitem{coch2009}
B.~Cochelin, C.~Vergez, {A} high order purely frequency-based
harmonic balance
  formulation for continuation of periodic solutions, Journal of Sound and
  Vibration 324~(1--2) (2009) 243--262.

\bibitem{laxa2009}
D.~Laxalde, F.~Thouverez, {C}omplex non-linear modal analysis for
mechanical
  systems {A}pplication to turbomachinery bladings with friction interfaces,
  Journal of Sound and Vibration 322~(4-5) (2009) 1009--1025.

\bibitem{krac2013a}
M.~Krack, L.~Panning-von Scheidt, J.~Wallaschek, {A} {M}ethod for
{N}onlinear
  {M}odal {A}nalysis and {S}ynthesis: {A}pplication to {H}armonically {F}orced
  and {S}elf-{E}xcited {M}echanical {S}ystems, accepted for publication in
  Journal of Sound and Vibration, doi:10.1016/j.jsv.2013.08.009.

\bibitem{krac2013d}
M.~Krack, L.~Panning-von Scheidt, J.~Wallaschek, A.~Hartung,
C.~Siewert,
  {R}educed {O}rder {M}odeling {B}ased on {C}omplex {N}onlinear {M}odal
  {A}nalysis and its {A}pplication to {B}laded {D}isks {W}ith {S}hroud
  {C}ontact, Paper GT2013-94560, Proceedings of ASME~Turbo Expo 2013, June 3-7,
  San Antonio, TX, USA (2013), 11pp.

\bibitem{blan2013}
F.~Blanc, C.~Touz{\'e}, J.-F. Mercier, K.~Ege, A.-S. Bonnet
Ben-Dhia, {O}n the
  numerical computation of nonlinear normal modes for reduced-order modelling
  of conservative vibratory systems, Mechanical Systems and Signal Processing
  36~(2) (2013) 520--539.

\bibitem{mane2001}
L.~I. Manevitch, {T}he {D}escription of {L}ocalized {N}ormal {M}odes
in a
  {C}hain of {N}onlinear {C}oupled {O}scillators {U}sing {C}omplex {V}ariables,
  Nonlinear Dynamics 25~(1-3) (2001) 95--109.

\bibitem{lee2005}
Y.~S. Lee, G.~Kerschen, A.~F. Vakakis, P.~Panagopoulos, L.~Bergman,
D.~M.
  McFarland, {C}omplicated dynamics of a linear oscillator with a light,
  essentially nonlinear attachment, Physica D: Nonlinear Phenomena 204~(1--2)
  (2005) 41--69.

\bibitem{vaka2008b}
A.~F. Vakakis, O.~V. Gendelman, G.~Kerschen, L.~A. Bergman, D.~M.
McFarland,
  Y.~S. Lee, {N}onlinear targeted energy transfer in mechanical and structural
  systems, Springer, 2008.

\bibitem{vaka2008}
A.~Vakakis, L.~Manevitch, Y.~Mikhlin, V.~Pilipchuk, A.~Zevin,
{N}ormal modes
  and localization in nonlinear systems, John Wiley {\&} Sons, 2008.

\bibitem{came1989}
T.~M. Cameron, J.~H. Griffin, {A}n {A}lternating {F}requency/{T}ime
{D}omain
  {M}ethod for {C}alculating the {S}teady-{S}tate {R}esponse of {N}onlinear
  {D}ynamic {S}ystems, Journal of Applied Mechanics 56~(1) (1989) 149--154.

\bibitem{guil1998}
J.~Guillen, C.~Pierre, {A}n {E}fficient, {H}ybrid,
{F}requency-{T}ime {D}omain
  {M}ethod for the {D}ynamics of {L}arge-{S}cale {D}ry-{F}riction {D}amped
  {S}tructural {S}ystems, Proc. of the IUTAM Symposium, August 3-7, Munich, Germany (1998), 1--10.

\bibitem{krac2013b}
M.~Krack, L.~Panning-von Scheidt, J.~Wallaschek, {A} {H}igh-{O}rder
{H}armonic
  {B}alance {M}ethod for {S}ystems {W}ith {D}istinct {S}tates, Journal of Sound
  and Vibration 332~(21) (2013) 5476--5488.

\bibitem{kers2009}
G.~Kerschen, M.~Peeters, J.~C. Golinval, A.~F. Vakakis, {N}onlinear
normal
  modes, {P}art {I}: {A} useful framework for the structural dynamicist:
  {S}pecial {I}ssue: {N}on-linear {S}tructural {D}ynamics, Mechanical Systems
  and Signal Processing 23~(1) (2009) 170--194.

\bibitem{groll2001b}
G.~v. Groll, D.~J. Ewins, {T}he harmonic balance method with
arc-length
  continuation in rotor/stator contact problems, Journal of Sound and Vibration
  241~(2) (2001) 223--233.

\bibitem{laza2010}
A.~Lazarus, O.~Thomas, {A} harmonic-based method for computing the
stability of
  periodic solutions of dynamical systems, Comptes Rendus M{\'e}canique 338~(9)
  (2010) 510--517.

\bibitem{sund1997}
P.~Sundararajan, S.~T. Noah, {D}ynamics of {F}orced {N}onlinear
{S}ystems
  {U}sing {S}hooting/{A}rc-{L}ength {C}ontinuation {M}ethod---{A}pplication to
  {R}otor {S}ystems, Journal of Vibration and Acoustics 119~(1) (1997) 9--20.

\bibitem{szem1979}
W.~Szemplinska-Stupnicka, {T}he modified single mode method in the
  investigations of the resonant vibrations of non-linear systems, Journal of
  Sound and Vibration 63~(4) (1979) 475--489.

\bibitem{nayf1979}
A.~H. Nayfeh, D.~T. Mook, {N}onlinear oscillations, John Wiley {\&}
Sons, New
  York 1979.

\bibitem{dahl1976}
P.~R. Dahl, {S}olid friction damping of mechanical vibrations, AIAA
Journal 14
  (1976) 1675--1682.

\bibitem{petr2012c}
E.~P. Petrov, {A}nalysis of {F}lutter-{I}nduced {L}imit {C}ycle
{O}scillations
  in {G}as-{T}urbine {S}tructures {W}ith {F}riction, {G}ap, and {O}ther
  {N}onlinear {C}ontact {I}nterfaces, Journal of Turbomachinery 134~(6) (2012)
  061018--061030.

\end{thebibliography}

\end{document}